\documentclass[aps,prx,twocolumn,secnumarabic,amsmath,amssymb,superscriptaddress,longbibliography]{revtex4-1}

\usepackage{latexsym}
\usepackage{amssymb}
\usepackage{amsmath}
\usepackage{amsfonts}
\usepackage{wasysym}
\usepackage{bm}

\usepackage{titlesec}
\usepackage{comment}
\usepackage{verbatim}

\usepackage{graphicx}
\usepackage{epsfig}
\usepackage{epstopdf}
\usepackage{subfigure}
\usepackage{tikz}
\usetikzlibrary{calc}

\usepackage{tabularx}
\usepackage{dcolumn}

\usepackage{booktabs}

\usepackage{color}
\usepackage[colorlinks,bookmarks=false,citecolor=blue,linkcolor=red,urlcolor=blue]{hyperref}

\definecolor{darkred}{rgb}{0.7,0.0,0.0}

\definecolor{darkblue}{rgb}{0,0.02,0.45}
\definecolor{darkgreen}{rgb}{0.02,0.45,0.0}

\definecolor{violet}{rgb}{0.8,0.2,0.6}

\def\be{\begin{equation}}
\def\ee{\end{equation}}
\def\bea{\begin{eqnarray}}
\def\eea{\end{eqnarray}}

\def\vec{\mathbf}
\def\bs{\boldsymbol}
\def\mc{\mathcal}

\makeatletter
\newcommand{\thickhline}{%
\noalign {\ifnum 0=`}\fi \hrule height 0.7pt
\futurelet \reserved@a \@xhline
}
\newcolumntype{"}{@{\hskip\tabcolsep\vrule width 0.7pt\hskip\tabcolsep}}
\makeatother

\usepackage{times}

\begin{document}
\date{\today}

\title{Phase diagram and quantum order by disorder in the Kitaev $K_1$-$K_2$ honeycomb magnet}

\author{Ioannis Rousochatzakis}
\affiliation{School of Physics and Astronomy, University of Minnesota, Minneapolis, MN 55455, USA}

\author{Johannes Reuther}
\affiliation{Dahlem Center for Complex Quantum Systems and Fachbereich Physik, Freie Universit\"at Berlin, 14195 Berlin, Germany}
\affiliation{Helmholtz-Zentrum Berlin f\"ur Materialien und Energie, 14109 Berlin, Germany}

\author{Ronny Thomale}
\affiliation{Institute for Theoretical Physics, University of W\"urzburg, 97074 W\"urzburg, Germany}

\author{Stephan Rachel}
\affiliation{Institute for Theoretical Physics, Technische Universit\"at Dresden, 01062 Dresden, Germany}

\author{N. B. Perkins}
\affiliation{School of Physics and Astronomy, University of Minnesota, Minneapolis, MN 55455, USA}

\begin{abstract}
We show that the topological Kitaev spin liquid on the honeycomb lattice is extremely fragile against the second neighbor Kitaev coupling $K_2$, which has been recently shown to be the dominant perturbation away from the nearest neighbor model in iridate Na$_2$IrO$_3$, and may also play a role in $\alpha$-RuCl$_3$ and Li$_2$IrO$_3$. This coupling explains naturally the zig-zag ordering (without introducing unrealistically large longer-range Heisenberg exchange terms), and the special entanglement between real and spin space observed recently in Na$_2$IrO$_3$. Moreover, the minimal $K_1$-$K_2$ model that we present here holds the unique property that the classical and quantum phase diagrams and their respective order by disorder mechanisms are qualitatively different due to the fundamentally different symmetries of the classical and quantum counterparts.
\end{abstract}

\maketitle

\pagebreak

\section{Introduction}\label{sec:intro}
The search for novel quantum states of matter arising from the interplay of strong electronic correlations, spin-orbit coupling (SOC), and crystal field splitting has recently  gained strong impetus in the context of $4d$ and $5d$ transition metal oxides~\cite{krempa14}. The layered iridates of the A$_2$IrO$_3$  (A=Na,Li) family~\cite{singh10,singh12,liu11,ye12,choi12,chun15} have been at the center of this search because of the prediction~\cite{jackeli09,jackeli10} that the dominant interactions in these magnets constitute the celebrated Kitaev model on the honeycomb lattice, one of the few exactly solvable models hosting gapped and gapless quantum spin liquids (QSLs)~\cite{kitaev06}. This aspect together with the realization that the Kitaev spin liquid is stable with respect to moderate Heisenberg-like perturbations~\cite{jackeli10,shaffer12} has triggered a lot of experimental activity on A$_2$IrO$_3$ and, more recently, on the similar $\alpha$-RuCl$_3$ compound~\cite{plumb14,sears15,kubota15}.

In the layered A$_2$IrO$_3$ magnets, the single-ion ground state configuration of Ir$^{4+}$ is an effective pseudospin $J_{\rm eff}\!=\!1/2$ doublet, where spin and orbital angular momenta are intertwined due to the strong SOC. In the original Kitaev-Heisenberg model proposed by Jackeli and Khaliullin~\cite{jackeli09}, the pseudospins couple via two competing nearest neighbor (NN) interactions: An isotropic antiferromagnetic (AFM) Heisenberg exchange, $J_1$, and a highly anisotropic Kitaev interaction, $K_1$, which is strong and ferromagnetic, a fact that is also confirmed by {\it ab-initio} quantum chemistry calculations by Katukuri {\it et al}~\cite{katukuri14,Satoshi2014}. Nevertheless, neither Na$_2$IrO$_3$ nor Li$_2$IrO$_3$ are found to be in the spin liquid state at low temperatures. Instead, they show, respectively, AFM zigzag and incommensurate long-range magnetic orders, none of which is actually present in the Kitaev-Heisenberg model for FM $K_1$ coupling.

\begin{figure}[!b] 
\includegraphics[width=0.6\columnwidth,angle=90,clip=true,trim=30 0 30 0]{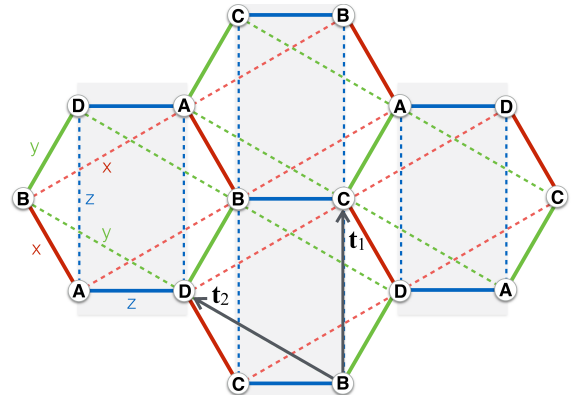}
\caption{(Color online) The Kitaev $K_1$-$K_2$ model with three types of NN (solid) and NNN (dashed) Ising bonds. Here $\vec{t}_1\!=\!a\vec{y}$ and $\vec{t}_2\!=\!(-\frac{\sqrt{3}}{2}\vec{x}\!+\!\frac{1}{2}\vec{y}) a$ are two primitive translations and $a$ is a lattice constant. We also show the vertical 2-leg ladders (shaded strips) discussed in the text, and the four-sublattice decomposition (A-D) related to the operations $H_{yzx}$ and $H_{xyz}$, see text.}\label{fig:Model}
\end{figure}

The most natural way to obtain these magnetic states is by including further neighbor Heisenberg couplings~\cite{Rastelli1979,Fouet2001,katukuri14,Satoshi2014}, which are non-negligible due to extended nature of the $5d$-orbitals of Ir$^{4+}$ ions~\cite{kimchi11,choi12}. In addition, recent calculations by Sizyuk {\it et al}~\cite{sizyuk14} based on the {\it ab-initio} density-functional data of Foyevtsova {\it et al}~\cite{katerina13} have shown that, for Na$_2$IrO$_3$, the next nearest neighbor (NNN) exchange paths must also give rise to an anisotropic, Kitaev-like coupling $K_2$,  which turns out to be AFM. More importantly, this coupling is the largest interaction after $K_1$. It has also been argued~\cite{rachel14} that $K_2$ plays an important role in the stabilization of the IC spiral state in Li$_2$IrO$_3$ and might be deduced from the strong-coupling limit of Hubbard model with topological band structure~\cite{shitade09,johannes12}.
 
Recent structural~\cite{plumb14} and magnetic~\cite{sears15} studies have shown that the layered honeycomb magnet $\alpha$-RuCl$_3$ is another example of a strong SOC Mott insulator, where the Ru$^{3+}$ ions are again described by effective $J_{\rm eff}\!=\!1/2$ doublets. At low $T$, this magnet exhibits zigzag ordering as in Na$_2$IrO$_3$. Furthermore, the superexchange derivations~\cite{shankar14,sizyuk15} based on the {\it ab initio} tight-binding parameters show that the NNN coupling $K_2$ is again appreciable, and the signs of both $K_1$ and $K_2$ are reversed compared to Na$_2$IrO$_3$ (i.e., $K_1$ is AFM and $K_2$ is FM). However, a strong off-diagonal symmetric NN exchange $\Gamma$ term~\cite{katukuri14,Satoshi2014,rau14}, which is allowed by symmetry, is also present~\cite{shankar14,sizyuk15}, together with a much smaller $J_1$ coupling. This compound must then be examined in connection to $\Gamma$, $K_2$, and $J_1$, since the $\Gamma$ term alone is not sufficient to explain the experimental situation, as we discuss at length in Sec.~\ref{sec:Disc}.

Motivated by these studies, here we consider the minimal extension of the NN Kitaev model that incorporates the effect of $K_2$, the $K_1$-$K_2$ model. We show that an extremely weak $K_2$ is enough to stabilize the zig-zag phases relevant for Na$_2$IrO$_3$ and $\alpha$-RuCl$_3$, without introducing  large, second and third neighbor Heisenberg exchange $J_2$ and $J_3$. While $J_2$ and $J_3$ are present in these compounds, the key point is that the Kitaev spin liquid is significantly more fragile against $K_2$ than $J_2$ and $J_3$. Thus, in conjunction with the above predictions from superexchange derivations, our findings suggest that any adequate minimal model of these compounds should include the NNN coupling $K_2$.

A very striking aspect of the zig-zag phases (shared by all magnetic phases) of the $K_1$-$K_2$ model is that they are only stabilized for quantum spins and not for classical spins, despite having a strong classical character. Indeed, these phases are Ising-like (with spins pointing along one of the three cubic axes), they are protected by a large excitation gap in the interacting $1/S$ spin-wave spectrum, and the spin lengths are extremely close to their classical value of $1/2$. Yet, these phases cannot be stabilized in the classical limit, in stark contrast to the  conventional situation where quantum and thermal fluctuations work in parallel and often lead to the same order-by-disorder phenomena. Instead, this rare situation we encounter here stems from the manifestly different symmetry structure of the classical and quantum Hamiltonians, and the underlying principle that time reversal can only act globally in quantum systems (see below). This aspect has important ramifications for the phase diagram at zero and finite temperatures $T$.

\section{Model \& Phase diagram}\label{sec:model}
The model we consider here is described by the effective spin-1/2 Hamiltonian 
\be\label{eq:Ham}
\mc{H}= K_1\sum_{\langle ij\rangle} S^{\gamma_{ij}}_iS_j^{\gamma_{ij}}+K_2 \sum_{\ll ij\gg} S^{\lambda_{ij}}_i S^{\lambda_{ij}}_j~,
\ee
where $\langle ij\rangle$ (respectively $\ll\!\!ij\!\!\gg$) label NN (NNN) spins on the honeycomb lattice, $S_j^a$ defines the $a$th cartesian component of the spin operator at site $j$, and $\gamma_{ij}$ ($\lambda_{ij}$) define the type of Ising coupling for the bond $(ij)$, see Fig.~\ref{fig:Model}. This model interpolates between two well known limits, the exactly solvable Kitaev spin liquid~\cite{kitaev06} at $K_2\!=\!0$, and the triangular Kitaev model at $K_1\!=\!0$~\cite{ioannis12,kimchi14,trebst14,jackeli15,Li2015}. It is easy to see that a finite $K_2$ ruins the exact solvability of the NN Kitaev model because the flux operators~\cite{kitaev06} $W_p\!=\!2^6S_1^zS_2^xS_3^yS_4^zS_5^xS_6^y$ (see site-labeling convention in Fig.~\ref{fig:PertTheory}, top left), around hexagons $p$ are no longer conserved. 

\begin{figure}[!t] 
\includegraphics[width=0.82\columnwidth,angle=90,clip=true,trim=0 0 0 0]{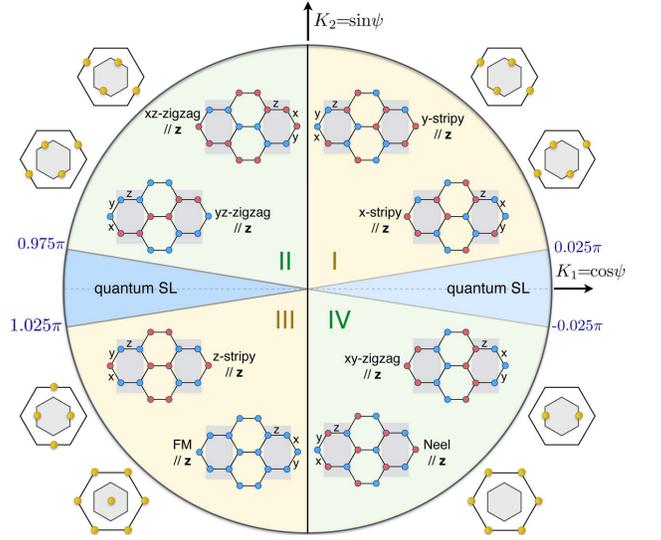}
\caption{(Color online) (a) The $T\!=\!0$ phase diagram of the model (\ref{eq:Ham}) as found by exact diagonalizations. Each of the magnetic regions (I-IV) hosts twelve degenerate quantum states. Here we show two members (where spins point along the $\vec{z}$-axis, blue/red circles denote spin up/down) that are related to each other by flipping the spins in every second ladder (shaded strips) of Fig.~\ref{fig:Model}. The Bragg peaks corresponding to $\langle S_i^zS_j^z\rangle$ correlations are also shown in the extended Brillouin zone (assuming the same magnetic form factor in the two unit cell sublattices). The corresponding Bragg reflections for $\langle S_i^xS_j^x\rangle$ and $\langle S_i^yS_j^y\rangle$ are related to $\langle S_i^zS_j^z\rangle$ by $\widetilde{\mathsf{C}}_{6\text{v}}$ spin-orbit rotations~\cite{SM}.} 
\label{fig:PhaseDiagram}
\end{figure}

In the following we parametrize $K_1\!=\!\cos\psi$ and $K_2\!=\!\sin\psi$, and take $\psi\!\in\![0,2\pi)$. It turns out that the physics actually remains the same under a simultaneous sign change of $K_1$ and $K_2$, because this can be gauged away by an operation $H_{yzx}\!=\!\prod_{i\in \text{B}}\mathsf{C}_{2y}(i)\prod_{j\in \text{C}}\mathsf{C}_{2z}(j)\prod_{k\in \text{D}}\mathsf{C}_{2x}(k)$, which is the product of $\pi$-rotations around the $\vec{y}$, $\vec{z}$, and $\vec{x}$ axis, respectively, for the B, C, and D sublattices of Fig.~\ref{fig:Model}. This {\it hidden} duality is a very common feature in many spin-orbital models~\cite{Khaliullin05,jackeli10,chaloupka15} but does not exist when Heisenberg couplings are also present (in contrast to the symmetry $H_{xyz}$ discussed below). Here it reduces our study to the first two quadrants of the unit circle of $\psi$.

\begin{figure*}[!t] 
\includegraphics[width=0.99\textwidth,angle=0,clip=true,trim=0 0 0 0]{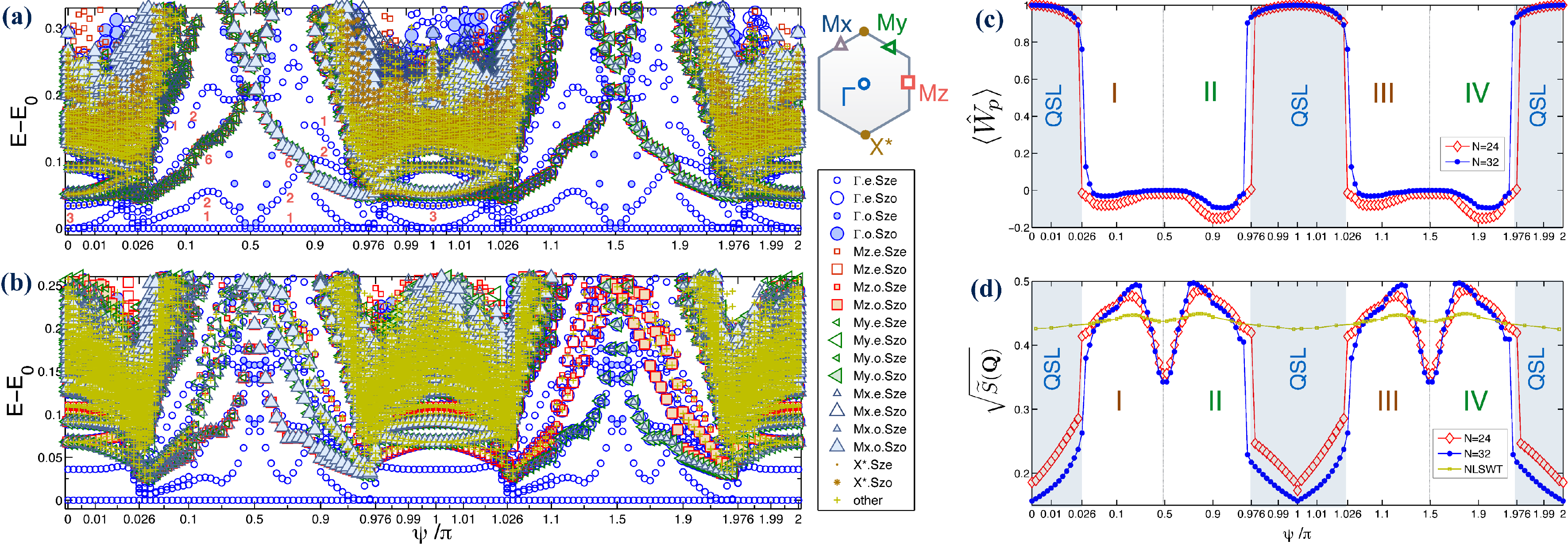}
\caption{(Color online) (a-b) Exact low-energy spectra (measured from the ground state energy $E_0$) of the 24-site (a) and 32-site (b) clusters, defined, respectively~\cite{SM}, by the spanning vectors $(2\vec{t}_1\!-\!4\vec{t}_2,4\vec{t}_1\!-\!2\vec{t}_2)$ and $(2\vec{t}_1\!-\!4\vec{t}_2,4\vec{t}_1)$. A  non-linear x-axis is used in order to highlight all regions of interest equally. The states are labeled by momenta $\vec{k}$ in the first BZ, parity (``e'' for even, ``o'' for odd) under inversion through hexagon centers, and parity under global spin $\pi$-rotations around the $\vec{x}$-axis (``Sze'' for even, ``Szo'' for odd). The (red) numbers in (a) denote the multiplicity of the lowest five levels in regions I and II, and the ground state degeneracy at $\psi\!=\!0$ and $\pi$. (c) Ground state expectation value $\langle W_p\rangle$ of Kitaev's flux operators. (d) Square root of the `symmetrized' ground state spin structure factor $\widetilde{\mc{S}}(\vec{Q})$ (see text), along with the spin length calculated from a self-consistent non-linear spin-wave theory (NLSWT).}\label{fig:ED}
\end{figure*}

Figure~\ref{fig:PhaseDiagram} shows the quantum phase diagram as found by exact diagonalizations (ED) on finite clusters, see discussion below and numerical data shown in Fig.~\ref{fig:ED}. There are six different regimes as a function of the angle $\psi$: the two quantum spin liquids (QSLs) regions (which have been enlarged for better visibility) around the exactly solvable Kitaev points ($\psi\!=\!0$ and $\pi$)  and four long-range magnetic regions (I-IV), hosting FM, Neel, stripy, as well as the zig-zag phases that are relevant for Na$_2$IrO$_3$ (II) and $\alpha$-RuCl$_3$ (IV). Under the duality transformation $H_{yzx}$, the two QSLs map to each other, I maps to III, and II maps to IV. 

Each of the magnetic regions actually hosts twelve degenerate quantum states, some of which are even qualitatively different among themselves, with very distinct Bragg reflections. For example, the region III hosts six FM and six stripy AFM ground states, and IV hosts six N\'eel and six zigzag AFM ground states. This striking aspect stems from a non-global symmetry, $H_{xyz}$, which is the product of $\pi$-rotations around the $\vec{x}$, $\vec{y}$, and $\vec{z}$ axis, respectively, for the B, C, and D sublattices of Fig.~\ref{fig:Model}. The two states shown in each magnetic region of Fig.~\ref{fig:PhaseDiagram} are related to each other by this symmetry, which for these particular states amounts to flipping the z-component of the spins in every second shaded ladder of Fig.~\ref{fig:Model}. The remaining ten states of the quantum ground state manifold arise by applying the global symmetries of the model: i) the double cover $\widetilde{\mathsf{C}}_{6\text{v}}$ of $\mathsf{C}_{6\text{v}}$, and ii) the double cover $\widetilde{\mathsf{D}}_2$ of the $\mathsf{D}_2$ group of global $\pi$ rotations in spin space.

Let us now turn to the numerical spectra shown in Fig.~\ref{fig:ED}~(a,b). First, the QSL regions are extremely narrow: They survive in a tiny window of $\delta\psi\!=\!0.05\pi$ around the exact Kitaev points, which is confirmed by the comparison of ED against large scale pseudofermion functional renormalization group (PFFRG) calculations~\cite{PhysRevB.81.144410,PhysRevB.83.024402,PhysRevB.84.014417,PhysRevB.84.100406}. So the QSLs are extremely fragile against $K_2$.

Second, Fig.~\ref{fig:ED}~(a,b) show very dense spectral features in the QSL regions, reflecting the continuum structure of fractionalized excitations above the Kitaev spin liquid. More specifically, for finite systems the ground state degeneracy at the exact Kitaev points~\footnote{This is a degeneracy between three out of the four topological sectors and can appear already for finite systems, depending on the cluster geometry and the corresponding structure of the boundary terms in the fermionic description of the problem~\cite{Kells09}.} is lifted by $K_2$. Still, for small enough $|K_2|$, the QSLs must be gapless in the thermodynamic limit, because $K_2$ respects time reversal symmetry and is therefore not expected~\cite{kitaev06} to open a gap in the Majorana spectrum~\footnote{However, a gap may eventually open at finite $K_2$, before the transitions to the magnetically ordered phases.}.

Third, unlike the QSL regions, the low-energy spectrum inside the magnetic regions is very discrete. In addition, most of the low-lying states within the energy window shown in Figs.~\ref{fig:ED}~(a,b) correspond precisely to the twelve quantum ground states discussed above. For finite systems, these states are admixed by a finite tunneling, leading to twelve symmetric eigenstates with quantum numbers corresponding to the decomposition of the symmetry broken states. This decomposition is worked out in detail in \cite{SM} and is indeed fully consistent with the ED data. So the lowest twelve states in each magnetic region of Figs.~\ref{fig:ED}~(a,b) will collapse to zero energy in the thermodynamic limit, leaving the true magnon excitations with a large anisotropy gap (modulo finite size corrections), reflecting the anisotropic, Ising-like character of the magnetic model.

Fourth, the magnetic instabilities, which serve as good examples of deconfinement-confinement transitions~\cite{Fradkin1979,Grignani1996,Tsuchiizu1999,mandal2011} for the underlying spinons, are of first order, as they are accompanied by finite, abrupt changes~\footnote{For finite systems, these are not true jumps because the transitions involve two states that belong to the same (identity) symmetry sector, leading to a very small level anticrossing.} in several ground state properties, e.g., in $\langle W_p\rangle$, and in the spin-spin correlations. Specifically, at $\psi\!=\!0$ and $\pi$, all fluxes $W_p$ have a value of $+1$~\cite{kitaev06}. A finite $K_2$ admixes sectors of different $W_p$, and so $\langle W_p\rangle$ drops continuously as we depart from the exact Kitaev's points, until it jumps to very low absolute values when we enter the magnetic phases, see Fig.~\ref{fig:ED}~(c). 

Turning to the spin-spin correlations, their abrupt change at the transition can be seen in the behavior of the `symmetrized' spin structure factor $\widetilde{\mc{S}}(\vec{Q})$ shown in Fig.~\ref{fig:ED}~(d), which is defined as 
\be
\widetilde{\mc{S}}(\vec{Q})\!=\!\frac{2}{N}\sum_{\alpha}\sum_{\vec{r}\neq 0} e^{i\vec{Q}^{(\alpha)}\cdot\vec{r}}\langle S^\alpha_0\! S^\alpha_{\vec{r}}\rangle~,
\ee
where $N$ is the number of sites, $\vec{Q}^{(\alpha)}$ is the ordering wavevector (see below) of the $\alpha$-th component of the spins ($\alpha\!=\!x,y,z$), and the extra factor of $2$ in this definition accounts for the fact~\cite{SM} that, for finite systems, there are no correlations between NN ladders like the ones shaded in Fig.~\ref{fig:Model}, due to the non-global symmetry $H_{xyz}$ discussed above. These data show clearly the short-range (long-range) character of spin-spin correlations inside (outside) the QSL regions. 

\begin{figure}[!t] 
\includegraphics[width=0.99\columnwidth,angle=0,clip=true,trim=0 0 0 0]{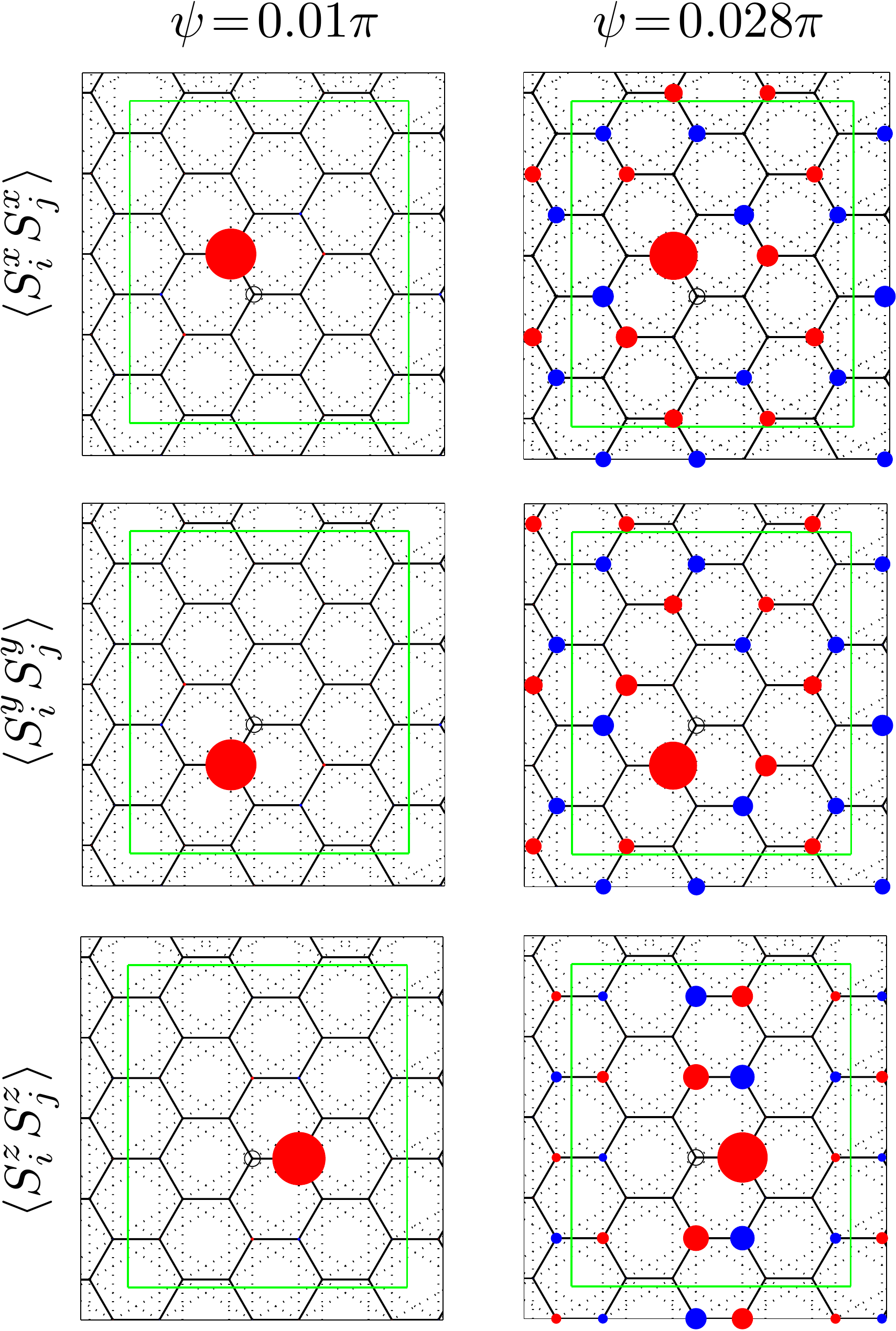}
\caption{Real-space spin-spin correlation profiles evaluated at the ground state of the $N\!=\!32$ cluster, inside the first QSL phase ($\psi\!=\!0.01\pi$, left column) and inside the magnetic phase I ($\psi\!=\!0.028\pi$, right column). Different rows correspond to the three different  channels $\langle S_i^\alpha S_j^\alpha\rangle$, $\alpha\!=\!x$, $y$ and $z$. The reference site $i$ is indicated by the small black open circle. Positive (negative) correlations are shown by filled blue (filled red) circles, whose radius scales with the magnitude of the correlation. The difference between $\alpha\!=\!z$ and $\alpha\!=\!x$, $y$ stems from the fact that the 32-site cluster does not have the full point-group symmetry of the infinite lattice, and the momentum point $\vec{M}_z$ is not equivalent by symmetry to $\vec{M}_{x}$ and $\vec{M}_y$, see~\cite{SM}.}\label{fig:EDLRSpinProfiles}
\end{figure}

This aspect can be seen more directly in Fig.~\ref{fig:EDLRSpinProfiles}, which shows the real-space spin-spin correlation profiles $\langle S_i^\alpha S_j^\alpha\rangle$, in the three channels $\alpha\!=\!x,y,z$, as calculated in the ground state of the 32-site cluster, inside the first QSL phase and slightly outside (magnetic phase I). The results show clearly the ultra short-range nature of the correlations inside the QSL region, and the long-range nature outside. 

Finally, the spin-spin correlation profiles demonstrate the special anisotropic character of the correlations, whereby different spin components $\alpha$ are correlated along different directions of the lattice (or, equivalently, different spin components $\alpha$ order at different ordering wavevectors $\vec{Q}^{(\alpha)}$, see also Fig.~\ref{fig:PhaseDiagram}), reflecting the locking between spin and orbital degrees of freedom in this model. Similar behavior is found for all other magnetic phases, including the zig-zag phases that are relevant for Na$_2$IrO$_3$ and $\alpha$-RuCl$_3$. Such a signature of directional dependent Kitaev couplings is exactly what has been reported recently by S.~H.~Chun {\it et al.} for Na$_2$IrO$_3$~\cite{chun15}; see also last paragraph of Sec.~\ref{sec:Disc}.

In the following we shall probe the physical mechanism of the spin liquid instabilities by taking one step back and examining the classical limit first.

\section{Classical limit}\label{sec:ClassLim}
For classical spins, the frustration introduced by the $K_2$ coupling is different from the one of the pure $K_1$ model studied by Baskaran {\it et al}~\cite{Baskaran08}. A straightforward classical minimization in momentum space~\cite{SM} gives lines of energy minima instead of a whole branch of minima~\cite{Baskaran08}, suggesting a sub-extensive ground state manifold structure, in analogy to compass-like models~\cite{Nussinov2015} or other special frustrated antiferromagnets~\cite{francisites}. 

We can construct one class of ground states by satisfying one of the three types of Ising bonds. We can choose for example the horizontal $zz$-bonds and align the spins along the $\vec{z}$-axis with relative orientations dictated by the signs of $K_1$ and $K_2$. The energy of the resulting configuration saturates the lower energy bound~\cite{SM} $E_b/(NS^2)\!=\!-|K_2|\!-\!|K_1|/2$ and is therefore one of the ground states. We can then generate other ground states by noting that $K_1$ and $K_2$ fix the relative signs of the spin projections $S_z$ only within the vertical 2-leg ladders of the lattice (shaded strips in Fig.~\ref{fig:Model}), but do not fix the relative orientation between different ladders, because these couple only via $xx$ and $yy$ Ising interactions which drop out at the mean field level. This freedom leads to $2^{n_{\text{lad}}}$ ground states, where $n_{\text{lad}}\!\propto\!\sqrt{N}$ is the number of vertical ladders. This sub-extensive degeneracy stems from the presence of non-global, {\it sliding} operations~\cite{Batista05,Nussinov2005,Nussinov2006,Nussinov2015} of flipping $S_z\!\mapsto\!-S_z$ for all spins belonging to one vertical ladder. Similarly, we can saturate the $xx$ or the $yy$ bonds, leading to 2-leg ladders running along the diagonal directions of the lattice. In total, this procedure delivers $3\!\times\!2^{n_{\text{lad}}}$ classical ground states. 

These states are actually connected in parameter space by valleys formed by other, continuous families of ground states that can be generated by global SO(3) rotations of the discrete states~\cite{SM}. The degeneracy associated with these valleys is accidental and can therefore be lifted by fluctuations. This is in fact the situation at finite $T$ where thermal fluctuations select one of the three types of discrete ground states, thereby breaking the three-fold symmetry of the model in the combined spin-orbit space. This corresponds to a finite-$T$ nematic phase where spins point along one of the three cubic axes but still sample all of the $2^{n_{\text{lad}}}$ corresponding states, without any long-range magnetic order. To achieve the latter one needs to spontaneously break all {\it sliding} symmetries and this cannot happen at finite $T$, according to the generalized Elitzur's theorem of Batista and Nussinov~\cite{Batista05}. The {\it sliding} symmetries can break spontaneously only at $T\!=\!0$ and in all possible ways, which is reflected in the divergence of the spin structure factor along lines in momentum space.

\section{Quantum spins \& Strong-coupling expansion}\label{sec:EffModel}
Turning to quantum spins, the situation is fundamentally different because the {\it sliding} symmetries are absent from the beginning: To flip one component of the spin we must combine a $\pi$-rotation in spin space and the time reversal operation~\footnote{By contrast, for the square-lattice compass model~\cite{Doucot2005,Dorier2005}, a $\pi$-rotation is actually enough (because the model involves only two types of Ising couplings), meaning that {\it sliding} symmetries exist also for quantum spins.}. The latter, however, involves the complex conjugation which cannot be constrained to act locally only on one ladder. Essentially, this means that the ladders must couple to each other dynamically by virtual quantum-mechanical processes, which in turn opens the possibility for long-range magnetic ordering even at finite $T$. 

\begin{figure}[!t] 
\includegraphics[width=0.75\columnwidth,angle=90,clip=true,trim=0 0 0 0]{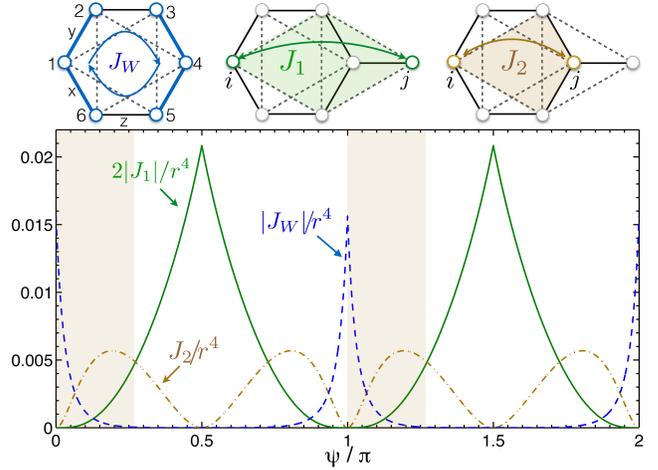}
\caption{(Color online) Top: The three types of virtual processes around the strong coupling limit $r\!=\!0$~\cite{SM}. Bottom: $|J_W|/r^4$, $2|J_1|/r^4$, and $J_2/r^4$ vs $\psi$. The shaded strips denote the regions where $J_2$ competes with $J_1$ and $J_2\!>\!2|J_1|$.}\label{fig:PertTheory}
\end{figure}

The natural way to understand the dynamical coupling between the ladders is to perform a perturbative expansion around one of the three strong coupling limits where the above discrete states become true quantum-mechanical ground states. Consider for example the limit where the $xx$ and $yy$ couplings, denoted by $K_1^{x(y)}$ and $K_2^{x(y)}$, are much smaller than the $zz$ couplings, $K_1^z$ and $K_2^z$. Let us also parametrize $K_{1,2}^{x(y)}\!=\!r K_{1,2}^z$, $K_1^z\!=\!\cos\psi$ and $K_2^z\!=\!\sin\psi$. For $r\!=\!0$ we have $n_{\text{lad}}$ decoupled vertical ladders, and $2^{n_{\text{lad}}}$ quantum ground states. Degenerate perturbation theory~\cite{SM} then shows that the degeneracy is first lifted at fourth order in $r$ via three, loop-four virtual processes that involve: (i) only $K_1^{x(y)}$, (ii) only $K_2^{x(y)}$, and (iii) both $K_1^{x(y)}$ and $K_2^{x(y)}$ perturbations, see the top panel of Fig.~\ref{fig:PertTheory}.

The processes (i) give rise to intra-ladder, six-body terms which are nothing else than the flux operators $W_p$. As shown by Kitaev~\cite{kitaev06}, these terms can be mapped to the square lattice Toric code~\cite{kitaev03} which has a gapped spin liquid ground state. Next, the processes (ii) and (iii) give rise to effective, NNN inter-ladder couplings of the form $J S_i^z S_j^z$, where $i$ and $j$ have the same (ii) or different (iii) sublattice unit cell indices, see top panel of Fig.~\ref{fig:PertTheory}. To fourth-order in $r$, the corresponding couplings $J_W$ (i), $J_{1}$ (ii), and $J_{2}$ (iii) read 
\small 
\bea
&&J_W\!=\!\frac{-\left(K_1^x K_1^y\right)^2 |K_1^z|}{64(|K_1^z|\!+\!2|K_2^z|)^2(|K_1^z|\!+\!3|K_2^z|)(|K_1^z|\!+\!4|K_2^z|)},\label{eq:Jw}\nonumber\\
&&J_{1}\!=\!\frac{\left(K_2^xK_2^y\right)^2}{8(|K_1^z|\!+\!2|K_2^z|)^2(2|K_1^z|\!+\!3|K_2^z|)}\text{sgn}(K_2^z),\label{eq:J1}\\
&&
J_{2}\!=\!\frac{K_1^x K_1^yK_2^x K_2^y}{4(|K_1^z|\!+\!2|K_2^z|)^3} \!\!\left[\!
\frac{|K_1^z|\!+\!|K_2^z|}{2|K_1^z|\!+\!3|K_2^z|} \!+\!\frac{2|K_2^z|}{|K_1^z|\!+\!4|K_2^z|}\!\right]\!\!.~~~\label{eq:J2}\nonumber
\eea 
\normalsize
Note that $J_{2}$ is always AFM and competes with $J_{1}$ in the regions I and III of Fig.~\ref{fig:PhaseDiagram}. We also emphasize that there is no $S_i^zS_j^z$ coupling when $i$ and $j$ belong to NN ladders. This is actually true to all orders in perturbation theory, because of the above non-global symmetry  $H_{xyz}$, which changes the sign of $S_z$ on every second vertical ladder (B and C sites of Fig.~\ref{fig:Model}).

The main panel of Fig.~\ref{fig:PertTheory} shows the behavior of $|J_W|/r^4$, $2|J_1|/r^4$, and $J_2/r^4$ as a function of the angle $\psi$, where the relative factor of $2$ between $|J_1|$ and $J_2$ accounts for their relative contribution to the total classical energy. Close to the exactly solvable points $\psi\!=\!0$ and $\pi$, the physics is dominated by the flux terms $W_p$ which, as mentioned above, lead to the gapped Toric code QSL~\cite{kitaev03,kitaev06}. The gapless QSL at $r\!=\!1$ is eventually stabilized by off-diagonal processes that necessarily admix states outside the lowest manifold of the $r\!=\!0$ point~\cite{Kai2008}.

The four magnetic phases I-IV of Fig.~\ref{fig:PhaseDiagram} are all stabilized by $J_1$ which, according to Fig.~\ref{fig:PertTheory}, is the dominant coupling in a wide region away from $\psi\!=\!0$ and $\pi$. Note that there are also two windows (shaded in Fig.~\ref{fig:PertTheory}) in the beginning of regions I and III where the two inter-ladder terms compete and $2|J_1|\!<\!J_2$. This opens the possibility for two more states (the ones favored by $J_2$) in these regions. This scenario is however not confirmed by our ED spectra and spin structure factors (especially for the 32-site cluster which is commensurate with both types of competing phases), showing that these phases are eventually preempted by the QSLs and the phases I and III at higher values of $r$.

We remark here that the 1-loop formulation of PFFRG delivers the $J_2$ but not the $J_{1}$ processes because, in a diagrammatic formulation of Abrikosov fermions, these processes relate to $3$-particle vertex contributions, which require a 2-loop formulation. However, for $\psi$ around $0$ and $\pi$, where $J_1$ is small, a 1-loop formulation already yields good agreement.

\section{Semiclassical picture}\label{sec:SeminClass}
The magnetic phases of the model can be captured by a standard semiclassical expansion, but this has to go beyond the non-interacting spin-wave level. Indeed, the zero-point energy of the quadratic theory lifts the accidental continuous degeneracy of the problem (selecting the cubic axes for the global direction in spin space, see Ref.~ [\onlinecite{SM}]), but fails to lift the discrete $2^{n_{\text{lad}}}$ degeneracy (the spectrum has lines of zero modes corresponding to the soft classical twists along individual ladders), and does not deliver a finite spin length, in analogy to several frustrated models~\cite{Khaliullin2001,Dorier2005,jackeli15,francisites}. The spurious zero modes are gapped out by spin-wave interactions, leading to the expected anisotropy gap and a finite spin length. The latter (obtained here from a self-consistent treatment of the quartic theory; details will be given elsewhere) tracks closely the behavior of the spin length extracted from the ED `symmetrized' spin structure factor~\footnote{The extra factor of $2$ in this definition accounts for the fact that there are no correlations between NN ladders for finite systems, due to the symmetry $H_{xyz}$, see also [\onlinecite{SM}]} $\widetilde{\mc{S}}(\vec{Q})$, see Fig.~\ref{fig:ED}~(d).  Furthermore, both methods give values that are very close to the classical value of $1/2$ inside the magnetic regions, showing that these phases are very robust. The quartic spin wave expansion is however insensitive to the proximity of the QSLs, most likely due to the first-order character of the transitions.

\section{Triangular Kitaev points}\label{sec:Tria} 
At $\psi\!=\!\pm\frac{\pi}{2}$ the system decomposes into two inter-penetrating triangular sublattices, where the $K_2$ coupling plays the role of a NN Kitaev coupling. This problem has been studied for both classical~\cite{ioannis12,kimchi14} and quantum spins~\cite{trebst14,jackeli15,Li2015}. The above analysis for the magnetic phases still holds here, the only difference being that the two legs of each ladder decouple, since they belong to different triangular sublattices. The ordering between the legs belonging to the same sublattice stems from the {\it effective} coupling $J_1$, which is the only one surviving at $K_1\!=\!0$. This coupling connects NNN legs only, leading to twelve states in each sublattice and thus $12^2$ states in total, instead of 12 for finite $K_1$. The accumulation of such extra states at low energies can be clearly seen in Fig.~\ref{fig:ED}(a-b) at $\psi\!=\!\pm\frac{\pi}{2}$. Note that while the ED spectra are broadly independent of system size, significant differences between the two cluster sizes are apparent near $\psi\!=\!\pm\pi/2$. These differences, e.g. on the ground state multiplicity, can be easily traced back to the different point group symmetry of the two clusters, see detailed explanation in \cite{SM}.

Finally we would like to point out that the origin of the ordering mechanism at the triangular Kitaev points has also been discussed independently in a recent paper by G.~Jackeli and A.~Avella~\cite{jackeli15}.

\section{Discussion}\label{sec:Disc}  
Charting out the stability region of the Kitaev spin liquid is an extremely relevant endeavor for the synthesis and characterization of new materials. One of the counterintuitive results of this study is that the frustrating (with respect to long-range magnetic order) NNN coupling $K_2$, which has exactly the same anisotropic form and symmetry structure as the $K_1$ term, destabilizes the Kitaev spin liquid much faster than the non-frustrating isotropic Heisenberg $J_1$ coupling. This finding gives a very useful hint in the search of realistic materials that exhibit the Kitaev spin liquid physics. In A$_2$IrO$_3$ materials, for example, the role of the size of the central ion (Na in Na$_2$IrO$_3$, or Li in Li$_2$IrO$_3$) in mediating the $K_2$ coupling (see also below) is a key aspect that can be easily controlled by experimentalists~\cite{Cao2013,Manni2014}.

On a more conceptual note, the physical mechanism underpinning the magnetic long range ordering in the present model is a novel example of order-by-disorder. Unlike many other classical states, here the ordering manifests only for quantum spins and not for classical spins. This striking contrast between classical and quantum spins is even more surprising in the light of the fact that all these phases have a strong classical character with local pseudo-spin lengths that are very close to the maximum classical value of $1/2$.

On this issue, we should stress that there is no discrepancy  between the very large pseudo-spin length that we report here and the small length of the magnetic moments extracted from magnetic reflections, e.g., in Na$_2$IrO$_3$~\cite{ye12}. Such an apparent discrepancy can be explained by the value of the $g$-factor which can be significantly smaller then $2$, because the orbital angular momentum is not quenched in strong SOC compounds. For the ideal cubic symmetry, for example, the well-known Land\'e formula gives $g=2/3$, and similar values could be expected for lower symmetry. 

Let us now elucidate further our main reasons on why the $K_2$ coupling must play an important role in Na$_2$IrO$_3$, and can be relevant in Li$_2$IrO$_3$ and $\alpha$-RuCl$_3$:

i) The super-exchange expansion of \cite{sizyuk14} shows clearly that the NNN Kitaev coupling is the second largest term in Na$_2$IrO$_3$, with $K_2\!\simeq\!7$-$9$~meV. All other perturbations are at most $1$-$2$~meV, consistent with the numbers given by the large-scale {\it ab initio} quantum chemistry study of \cite{katukuri14}. The mechanism behind the large magnitude of $K_2$ in Na$_2$IrO$_3$ is physically very clear: It originates from the large diffusive Na ions that reside in the middle of the exchange pathways, and the constructive interference of a large number of four pathways~\cite{sizyuk14}.

In Li$_2$IrO$_3$, the $K_2$ interaction comes from the same mechanism but it is relatively smaller because of the smaller size of Li ions~\cite{sizyuk15}. Still, as discussed in \cite{rachel14}, this coupling can be important to explain the current experimental evidence in terms of magnetic susceptibility profile, Curie-Weiss temperature, and the relevant range of couplings. 

Finally, in $\alpha$-RuCl$_3$, the analogous super-exchange path is absent, but an appreciable $K_2$ still arises from the anisotropy of diagonal interactions originated from the interplay between different hopping processes~\cite{sizyuk15}. However, as we already pointed out in the Introduction, the second largest coupling in $\alpha$-RuCl$_3$ is the anisotropic exchange $\Gamma$~\cite{katukuri14,rau14}. According to the study of J.~Rau {\it et al.}~\cite{rau14}, a positive $\Gamma$ seems to compete with $K_2$ for positive $K_1$~\cite{sizyuk15}. However, the situation is still unclear since the Bragg peaks of the states favored by $\Gamma$ do not reside at the $\vec{M}$ points of the BZ found experimentally by J. A. Sears {\it et al.}~\cite{sears15}, whereas such Bragg peaks are naturally present in the zig-zag phases favored by $K_2$, or even by a negative $J_1$.  So a lot more work is needed to clarify the relative importance of $\Gamma$, $K_2$, and $J_1$ in $\alpha$-RuCl$_3$.
 
ii) The $K_2$ coupling explains naturally the zig-zag ordering in Na$_2$IrO$_3$. This phase cannot arise in the original $J_1$-$K_1$ model, because this would require an AFM coupling $K_1$, whereas it is widely accepted that $K_1$ is FM and large in magnitude, see e.g. \cite{Satoshi2014}. Also, the much smaller $\Gamma$ terms, which are positive, also favor the zig-zag phase and do not compete with $K_2$, according to \cite{rau14}.

iii) The $K_2$ coupling can provide in addition the basis to resolve the long-standing puzzle of the large AFM Curie-Weiss temperature~\cite{singh10,singh12,choi12}, without incorporating unrealistically large values of longer-range Heisenberg couplings $J_2$ and $J_3$. 

iii) The recent diffusive x-ray scattering experiments by S.~H.~Chun {\it et al.}~\cite{chun15} have provided direct evidence for the predominant role of anisotropic, bond directional interactions in Na$_2$IrO$_3$. In conjunction with the above discussion and the results of Fig.~\ref{fig:EDLRSpinProfiles}, the $K_2$ term then emerges naturally as the number one anisotropic candidate term that can drive the zig-zag ordering and the directional dependence of the scattering found in \cite{chun15}.

An aspect that remains to be discussed in the context of Na$_2$IrO$_3$ is the direction of the magnetic moments which, according to the x-ray scattering data of S.~H.~Chun {\it et al.}~\cite{chun15}, do not point along the cubic axes but along the face diagonals. As discussed above, the K$_2$ coupling stabilizes the zig-zag phase but it is unable to lock the direction of the moments at the mean-field level due to an infinite accidental degeneracy. The fact that the locking along the cubic axes in the K$_1$-K$_2$ model eventually proceeds via a quantum order-by-disorder process (see Ref.~[\onlinecite{SM}]) renders this result very susceptible to much smaller anisotropic interactions that can pin the direction of the moments already at the mean field level. A very small positive anisotropic $\Gamma$ term can for example play such a role and can account for the locking along the face diagonals, as can be directly seen by a straightforward minimization of the classical energy. An alternative scenario involves a competing order-by-disorder effect within a more extended model that includes weak longer-range exchange interactions~\cite{sizyuk15}.

{\it Acknowledgements.}
We acknowledge the Minnesota Supercomputing Institute (MSI) at the University of Minnesota and the Max Planck Institute for the Physics of Complex Systems, Dresden, where a large part of the numerical computations took place. We are also grateful to R. Moessner, C. Price, O. Starykh, G. Jackeli, Y. Sizyuk, P. Mellado, and M. Schulz for stimulating discussions. I.R. and N.B.P. acknowledge the support from NSF Grant DMR-1511768. 
J.R. was supported by the Frei Universit\"at Berlin within the Excellence Initiative of the German Research Foundation. 
R.T. was supported by the European Research Council through ERC-StG-336012 and by DFG-SFB 1170. S.R. was supported by DFG-SFB 1143, DFG-SPP 1666, and by the Helmholtz association through VI-521. 
S.R., R.T. and N.B.P.  acknowledge the hospitality of the KITP during the program ``New Phases and Emergent Phenomena in Correlated Materials with Strong Spin-Orbit Coupling'' and a partial support by the National Science Foundation under grant No. NSF PHY11-25915.


%

\clearpage

\appendix

\pagenumbering{roman}

\begin{widetext}
\begin{center}
\Large{{\bf Supplemental material}}
\end{center}

\setcounter{figure}{0}







In this Supplementing material we provide auxiliary information and technical details and derivations. Specifically, Sec.~\ref{Sec:SM} deals with the Luttinger-Tisza minimization of the classical energy in momentum space (\ref{Sec:LT}), and the order-by-disorder process by harmonic spin-waves (\ref{Sec:LSWT}). Sec.~\ref{Sec:ED} gives details about our finite-size ED study, including the symmetry analysis of the low-energy spectra in regions I and II of the phase diagram (\ref{Sec:Spectr}), and the definition of the `symmetrized' spin structure factor $\widetilde{\mc{S}}(\vec{Q})$. In Sec.~\ref{PFFRG} we provide results from the pseudofermion functional renormalization group (PFFRG) approach. Finally, in Sec.~\ref{Sec:Heff} we provide the derivation of the effective Hamiltonian around the strong coupling limit of $K_{1,2}^{x(y)}\!=\!0$.

\section{Semiclassical analysis}\label{Sec:SM}
\subsection{Lutinger-Tisza minimization}\label{Sec:LT}
\begin{figure}[!b] 
\includegraphics[width=0.375\columnwidth,angle=0,clip=true,trim=0 0 0 0]{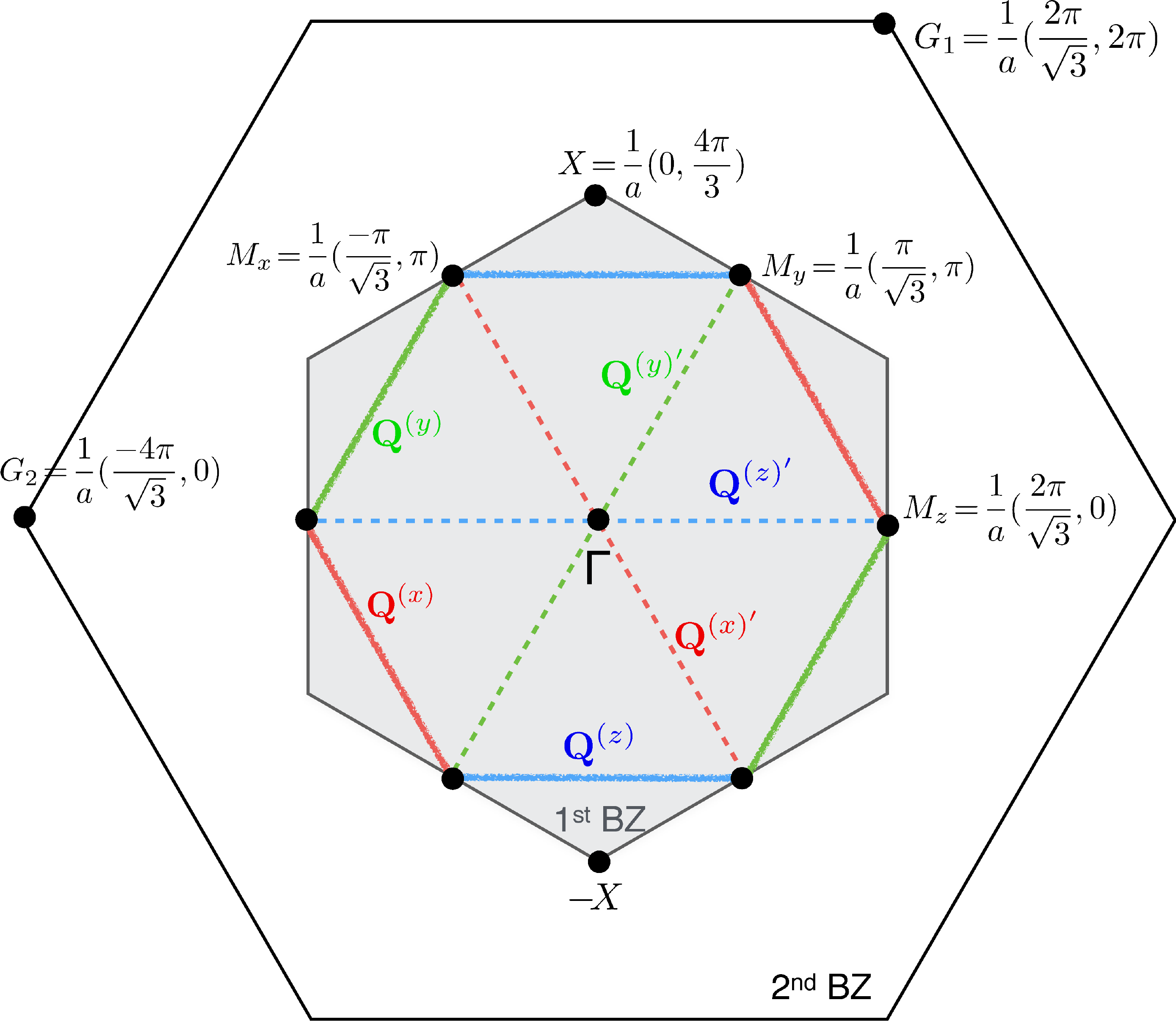}
\caption{The first two Brilouin zones of the honeycomb lattice, along with the special lines in momentum space $\vec{Q}^{(x)}$, $\vec{Q}^{(y)}$, and $\vec{Q}^{(z)}$ (respectively $\vec{Q}^{(x)'}$, $\vec{Q}^{(y)'}$, and $\vec{Q}^{(z)'}$) corresponding to the minima of the classical energy for $K_2\!>\!0$ ($<0$), see text.}\label{SMfig:BZ}
\end{figure}
We choose the primitive vectors of the honeycomb lattice as $\vec{t}_1\!=\!a\vec{y}$ and $\vec{t}_2\!=\!(-\frac{\sqrt{3}}{2}\vec{x}\!+\!\frac{1}{2}\vec{y}) a$, where $a$ is a lattice constant, see Fig.~1 of the main paper.  We also define $\vec{t}_3\!=\!\vec{t}_1\!-\!\vec{t}_2\!=\!\frac{\sqrt{3}}{2}\vec{x}\!+\!\frac{1}{2}\vec{y}$. In the following, we label the Bravais lattice vectors as $\vec{R}\!=\!n\vec{t}_1\!+\!m\vec{t}_2$, where $n$ and $m$ are integers. We also denote the two sites in the unit cell by a sublattice index $i\!=\!1$-$2$. The total classical energy of the $K_1$-$K_2$ model reads
\small
\bea
E = \sum_{\vec{R}} K_1 \left( 
S_{\vec{R},1}^z~S_{\vec{R},2}^z
+S_{\vec{R},1}^x~S_{\vec{R}+\vec{t}_2,2}^x
+S_{\vec{R},1}^y~S_{\vec{R}-\vec{t}_3,2}^y
\right)
+K_2 ~\sum_{\vec{R},i}\left( 
S_{\vec{R},i}^z~S_{\vec{R}-\vec{t}_1,i}^z
+S_{\vec{R},i}^x~S_{\vec{R}+\vec{t}_3,i}^x
+S_{\vec{R},i}^y~S_{\vec{R}+\vec{t}_2,i}^y
\right)~.
\eea
\normalsize
Defining $\vec{S}_{\vec{R},i} \!=\! \sum_{\vec{k}} e^{i\vec{k}\cdot \vec{R}}~\vec{S}_{\vec{k},i}$, we get
\small
\bea
\epsilon \equiv  \mc{H}/N_{uc} &=& K_1 \sum_{\vec{k}}  \left[
S_{\vec{k},1}^z~S_{-\vec{k},2}^z
+e^{-i \vec{k}\cdot \vec{t}_2}S_{\vec{k},1}^x~S_{-\vec{k},2}^x
+e^{i \vec{k}\cdot \vec{t}_3}S_{\vec{k},1}^y~S_{-\vec{k},2}^y
 \right] \nonumber\\
&+&K_2 \sum_{\vec{k},i}\left[
\cos(\vec{k} \cdot \vec{t}_1)~ S_{\vec{k},i}^z~S_{-\vec{k},i}^z
+\cos(\vec{k}\cdot \vec{t}_3)~ S_{\vec{k},i}^x~S_{-\vec{k},i}^x
+\cos(\vec{k}\cdot \vec{t}_2)~ S_{\vec{k},i}^y~S_{-\vec{k},i}^y
 \right]\nonumber\\
&=&\sum_{\vec{k},ij}\sum_{\alpha} S_{\vec{k},i}^\alpha \cdot \Lambda^{(\alpha)}_{ij}(\vec{k}) \cdot S_{-\vec{k},j}^\alpha~,\nonumber
\eea
\normalsize
where $N_{uc}\!=\!N/2$, is the number of unit cells, and the matrices $\bs{\Lambda}^{(\alpha)}$ (where $\alpha\!=\!x,y,z$) are given by
\small
\bea
\bs{\Lambda}^{(x)}(\vec{k}) \!\!=\!\! \left(\!\!
\begin{array}{ll}
K_2 \cos(\vec{k}\cdot\vec{t}_3) & \frac{K_1}{2} e^{-i\vec{k}\cdot\vec{t}_2} \\
\frac{K_1}{2} e^{i\vec{k}\cdot\vec{t}_2} & K_2 \cos(\vec{k}\cdot\vec{t}_3)
\end{array}
\!\!\right)\!\!,~~
\bs{\Lambda}^{(y)}(\vec{k}) \!\!=\!\! \left(\!\!
\begin{array}{ll}
K_2 \cos(\vec{k}\cdot\vec{t}_2) & \frac{K_1}{2} e^{i\vec{k}\cdot\vec{t}_3} \\
\frac{K_1}{2} e^{-i\vec{k}\cdot\vec{t}_3} & K_2 \cos(\vec{k}\cdot\vec{t}_2)
\end{array}
\!\!\right)\!\!,~~
\bs{\Lambda}^{(z)}(\vec{k}) \!\!=\!\! \left(\!\!
\begin{array}{ll}
K_2 \cos(\vec{k}\cdot\vec{t}_1) & \frac{K_1}{2} \\
\frac{K_1}{2} & K_2 \cos(\vec{k}\cdot\vec{t}_1)
\end{array}
\!\!\right)\!\!.\nonumber
\eea
\normalsize
To find the classical minimum we need to minimize the energy under the strong constraints $\vec{S}_{\vec{R},i}^2 \!=\!S^2$, $\forall (\vec{R}, i)$. The Luttinger-Tisza method~\cite{LTSM, BertautSM, LitvinSM, KaplanSM} amounts to relax the strong constraints with the weaker one  $\sum_{\vec{R},i} \vec{S}_{\vec{R},i}^2 \!=\! N S^2$, or equivalently $\sum_{\vec{k},i} \vec{S}_{\vec{k},i}\cdot \vec{S}_{-\vec{k},i} \!=\!S^2$. If we can find a minimum under the weak constraint that also satisfies the strong constraints then we have solved the problem. To this end, we minimize the function 
\small
\be
F = \epsilon - \lambda \sum_{\vec{k},i} (\vec{S}_{\vec{k},i}\cdot \vec{S}_{-\vec{k},i}  - S^2)~,
\ee
\normalsize
with respect to $\{S_{-\vec{k},i}^\alpha\}$, which gives a set of three eigenvalue problems for the $\bs{\Lambda}$ matrices:
\small
\be
\sum_{j=1,2}\Lambda^{(\alpha)}_{ij}(-\vec{q}) ~ S_{\vec{q},j}^\alpha = \lambda ~ S_{\vec{q},i}^\alpha,~~\alpha=x, y, z~.
\ee
\normalsize
If we can satisfy these three relations (plus the strong constraint) with a single eigenvalue $\lambda$, then $\epsilon\!=\!\lambda S^2$. So the energy minimum corresponds to the minimum over the three eigenvalues $\lambda^{(\alpha)}$ of the matrices $\bs{\Lambda}^{(\alpha)}(-\vec{k})$, and over the whole Brillouin zone (BZ). 
The eigenvalues of these matrices and the corresponding eigenvectros are:
\small
\bea
&&
\lambda^{(x)}_\pm=K_2 \cos(\vec{k}\cdot\vec{t}_3)\pm \frac{1}{2}K_1,~~~
\lambda^{(y)}_\pm=K_2 \cos(\vec{k}\cdot\vec{t}_2)\pm \frac{1}{2}K_1,~~~
\lambda^{(z)}_\pm=K_2 \cos(\vec{k}\cdot\vec{t}_1)\pm \frac{1}{2}K_1,\nonumber\\
&&
\vec{v}^{(x)}_\pm \sim \left(\begin{array}{c}
1 \\
\pm e^{i\vec{k}\cdot\vec{t}_2}
\end{array}\right),~~~ 
\vec{v}^{(y)}_\pm \sim \left(\begin{array}{c}
1 \\
\pm e^{-i\vec{k}\cdot\vec{t}_3}
\end{array}\right),~~~ 
\vec{v}^{(z)}_\pm \sim \left(\begin{array}{c}
1 \\
\pm 1
\end{array}\right)~.\nonumber
\eea
\normalsize
For $K_2$ positive, the minima of $\lambda^{(x)}_\pm$, $\lambda^{(y)}_\pm$, and $\lambda^{(z)}_\pm$ are located on the lines $\vec{Q}^{(x)} \!=\! r( \vec{G}_1\!+\!\vec{G}_2) \!+\! (l\!+\!\frac{1}{2})\vec{G}_2$, $\vec{Q}^{(y)}\!=\! r \vec{G}_1 \!+\! (l\!+\!\frac{1}{2})\vec{G}_2$, and $\vec{Q}^{(z)} \!=\! r \vec{G}_2 \!+\! (l\!+\!\frac{1}{2})\vec{G}_1$, respectively, where $l$ is any integer and $r\in(-\frac{1}{2},\frac{1}{2})$. On the other hand, for $K_2$ negative, the minima are located on the lines: $\vec{Q}^{(x)'} \!=\! r( \vec{G}_1\!+\!\vec{G}_2) \!+\! l\vec{G}_2$, $\vec{Q}^{(y)'}\!=\! r \vec{G}_1 \!+\! l\vec{G}_2$, and $\vec{Q}^{(z)'} \!=\! r \vec{G}_2 \!+\! l \vec{G}_1$. Both sets of lines are shown in Fig.~\ref{SMfig:BZ}.

Let us now try to build a ground state from the minima of the above eigenvectors for the case $K_{1,2}\!>\!0$, by using the line of minima $\vec{Q}^{(z)}$ as follows:
\small
\be
\left(\begin{array}{c}
S^z_{\vec{R},1}\\
S^z_{\vec{R},2}
\end{array}\right) 
=S\sum_{\{\vec{Q}^{(z)}\}} f_{\vec{Q}^{(z)}} e^{i \vec{Q}^{(z)}\cdot\vec{R}} 
\left(\begin{array}{c}
1\\
-1
\end{array}\right)
= 
(-1)^n S \left(\begin{array}{c}
\xi_m\\
-\xi_m
\end{array}\right)~,
\ee
\normalsize
where we used the relation $\vec{R}\!=\!n\vec{t}_1+m\vec{t}_2$ and have defined $\xi_m \equiv \int_{-1/2}^{1/2} dr f(r) e^{i 2\pi m r}$, which is the Fourier transform of the envelope function $f(r)$. We still need to satisfy the spin length constraint, which imposes a condition that the inverse Fourier transform of $f(r)$ takes only the values $\pm 1$. This freedom corresponds to the sliding symmetries of flipping individual vertical ladders, and leads to $2^{n_{\text{lad}}}$ degenerate states (where $n_{\text{lad}}$ is the number of vertical ladders), as discussed in the main text.

Similarly we can construct another $2\times 2^{n_{\text{lad}}}$ states by using the lines $\vec{Q}^{(x)}$ or $\vec{Q}^{(y)}$ in momentum space, which correspond to decoupled ladders running along the diagonal directions of the lattice. Altogether, we have found the $3\times 2^{n_{\text{lad}}}$ discrete classical ground states discussed in the main text by using the Luttinger-Tisza minimization method. 

Finally, it is easy to see that we can also combine the three types of states into a continuous family of other ground states that include coplanar and non-coplanar states. This family can be parametrized by two angles $\theta$ and $\phi$ as follows, 
\be
\vec{S}_{\vec{R},i} = S\left( \sin\theta\cos\phi ~S_{\vec{R},i}^x\vec{x} 
+\sin\theta\sin\phi ~S_{\vec{R},i}^y\vec{y}
+\cos\theta~S_{\vec{R},i}^z\vec{z} \right),
\ee
where $i=1,2$ and $S_{\vec{R},i}^x$, $S_{\vec{R},i}^y$ and $S_{\vec{R},i}^z$ denote the three type of discrete solutions discussed above.

\subsection{Harmonic order-by-disorder}\label{Sec:LSWT}
\begin{figure}[!t] 
\includegraphics[width=0.9\columnwidth,angle=0,clip=true,trim=0 0 0 0]{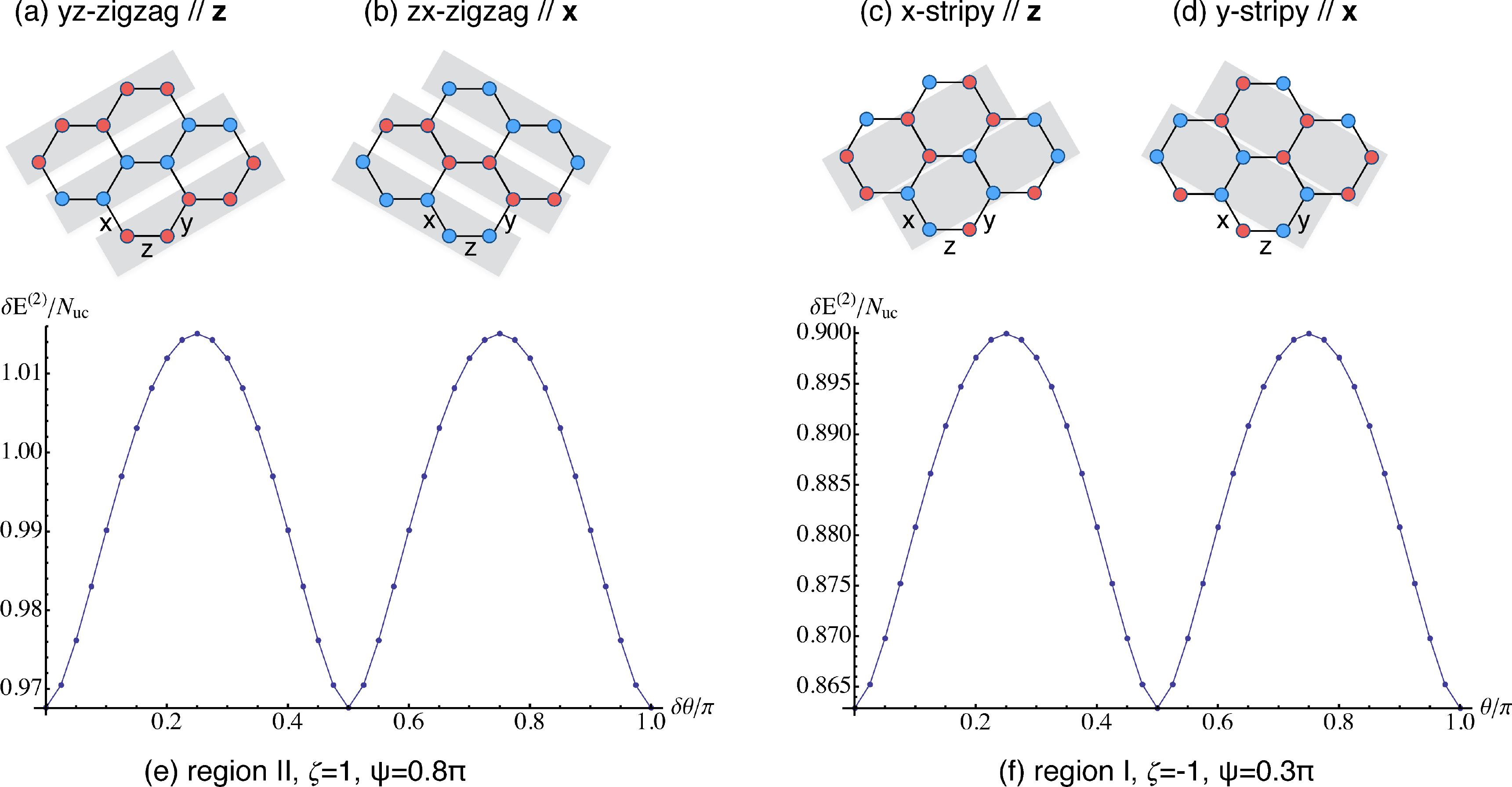}
\caption{Two representative zigzag (a-b) and two stripy (c-d) phases that belong to the classical ground state manifold inside the regions II and I, respectively. Blue (red) circles denote spins pointing up (down) along the $\vec{z}$-axis for (a,c) or along the $\vec{x}$-axis for (b,d). The shaded stripes denote the FM zigzag lines in each of the two zigzag phases (a-b), or the FM ladders in each of the two stripy phases (c-d).
(e-f) Harmonic zero-point energy $\delta E^{(2)}$ (divided by the number of unit cells $N_{uc}$) as a function of the parameter $\theta$, for two representative points inside region II (e) and I (f).}\label{SMfig:2str2zz}
\end{figure}
As we claimed in the main text, harmonic spin waves lift the accidental continuous degeneracy of the classical ground state manifold and select the discrete $3\times2^{n_{\text{lad}}}$ states, whereby spins point along the cubic axes. Here we shall demonstrate this result by considering a one-parameter family of coplanar states obtained by linearly combining two zigzag states and two stripy states with spins pointing along the cubic axes. In the  resulting family of states, spins are pointing in some direction on the {\bf zx}-plane. 

Figure~\ref{SMfig:2str2zz} shows the two zigzag and two stripy phases with spins pointing along the cubic axes. 
Here ``yz-zigzag//{\bf x}'' denotes a zigzag state with FM zig-zag lines running along the yy and zz bonds of the Kitaev Hamiltonian, and the spins point along the ${\bf x}$-axis. 
Similarly, ``x-stripy//{\bf z}'' denotes a stripy state with FM ladders formed by the xx bonds of the Kitaev Hamiltonian, and the spins point along the ${\bf z}$-axis. 
Specifically, these states can be written as:
\small
\bea
\left(\begin{array}{c}
\vec{S}_{\vec{R},1}\\ 
\vec{S}_{\vec{R},2}
\end{array}\right) 
&=&  e^{i \vec{M}_x\cdot\vec{R}} 
\left(\begin{array}{c}
\vec{z}\\
\vec{z}
\end{array}\right)
= (-1)^{n+m}
\left(\begin{array}{c}
\vec{z}\\
\vec{z}
\end{array}\right)
 ~~
\to yz\text{-zigzag //}  {\bf z} 
\nonumber\\
&=& e^{i \vec{M}_y\cdot\vec{R}} 
\left(\begin{array}{c}
\vec{x}\\
\vec{x}
\end{array}\right)
= (-1)^n 
\left(\begin{array}{c}
\vec{x}\\
\vec{x}
\end{array}\right)
\to zx\text{-zigzag //}  {\bf x} \nonumber\\
&=&
e^{i \vec{M}_x\cdot\vec{R}} 
\left(\begin{array}{c}
\vec{z}\\
-\vec{z}
\end{array}\right) 
= (-1)^{n+m} 
\left(\begin{array}{c}
\vec{z}\\
-\vec{z}
\end{array}\right) 
\to x\text{-stripy //} {\bf z} \nonumber\\
&=& 
e^{i \vec{M}_y\cdot\vec{R}} 
\left(\begin{array}{c}
\vec{x}\\
-\vec{x}
\end{array}\right)
= (-1)^n 
\left(\begin{array}{c}
\vec{x}\\
-\vec{x}
\end{array}\right)
\to y\text{-stripy //} {\bf x} \nonumber
\eea
\normalsize
where $\vec{M}_x=\left(-\frac{\pi}{\sqrt{3}},\pi\right)$ and $\vec{M}_y=\left(\frac{\pi}{\sqrt{3}},\pi\right)$ (see Fig.~\ref{SMfig:BZ}) and $\vec{R}=n\vec{t}_1+m\vec{t}_2$.
The one-parameter family of classical ground states are obtained by linear combinations of the above states: 
\small
\bea
\left(\begin{array}{c}
\vec{S}_{\vec{R},1}\\ 
\vec{S}_{\vec{R},2}
\end{array}\right) =
(-1)^n
\left(\begin{array}{c}
(-1)^m\cos\theta \vec{z} + \sin\theta \vec{x}\\
\zeta(-1)^m\cos\theta \vec{z} +\zeta \sin\theta \vec{x}
\end{array}\right),
\eea
\normalsize
where 
$\zeta=1$ for the zigzag case and $\zeta=-1$ for the stripy case.
The effect of harmonic spin waves can be found by a standard linear spin-wave expansion around the corresponding states for each value of $\theta$. 
Figs.~\ref{SMfig:2str2zz}~(e-f) show the zero-point energy correction (per number of unit cells) as a function of the angle $\theta$ for a representative point inside region II ($\psi=0.8\pi$, $\zeta=1$) and another point inside region I ($\psi=0.3\pi$, $\zeta=-1$).  
The data show clearly that harmonic fluctuations select the states with the spins pointing along the cubic axes ($\theta=0$, $\pm \pi/2$, and $\pi$).

We have checked that the result is the same for the corresponding order-by-disorder process for the one-parameter family of states obtained by combining two states with the same wavevector, such as the ``zx-zigzag // {\bf z}'' and ``zx-zigzag // {\bf x}''.

\section{Technical details about the ED study}\label{Sec:ED}

\begin{figure}[!b] 
\includegraphics[width=0.33\columnwidth,angle=90,clip=true,trim=180 0 180 0]{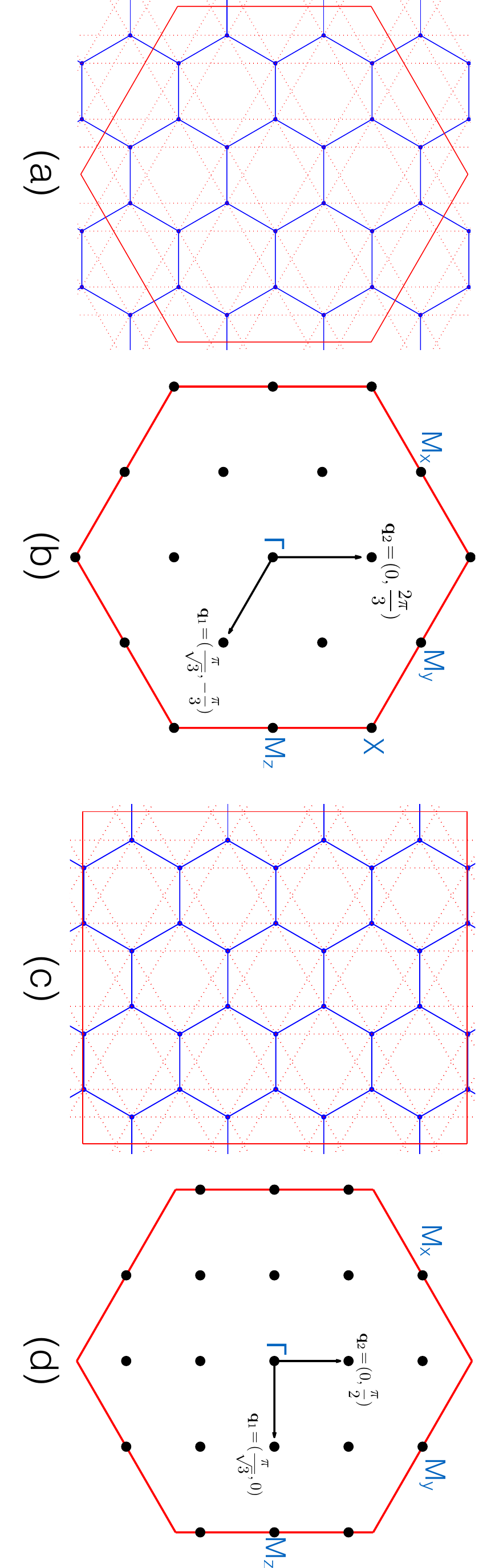}
\caption{The two finite clusters (with periodic boundary conditions) used in our ED study, with $N\!=\!24$ (a) and $32$ (c) sites, along with the allowed momenta in the first Brillouin zone, (b) and (d), respectively. }\label{SMfig:EDclusters}
\end{figure}

\subsection{The symmetry group of the Hamiltonian}
The full symmetry group of the $K_1$-$K_2$ model, for half-integer spins, is $\mc{T}\times \widetilde{C}_{6\text{v}}\times\widetilde{\mathsf{D}}_2$, which consists of: 

\begin{enumerate}
\item The translation group $\mc{T}$ generated by the primitive translation vectors $\vec{t}_1$ and $\vec{t}_2$, see Fig.~1 of the main text.  

\item The double cover $\widetilde{\mathsf{C}}_{6\text{v}}$ of the group $\mathsf{C}_{6\text{v}}\subset\mathsf{SO}(3)$ in the combined spin and real space, where the six-fold axis goes through one of hexagon centers. This group is generated by two operations: the six-fold rotation $\mathsf{C}_6$ around $[111]$, whose spin part maps the components $(x,y,z)\mapsto (y,z,x)$, and the reflection plane $(1\bar{1}0)$ that passes through the $zz$-bonds of the model, whose spin part maps $(x,y,z)\mapsto (-y,-x,-z)$. 

\item The double cover $\widetilde{\mathsf{D}}_2$ of the point group $\mathsf{D}_2\subset\mathsf{SO}(3)$, which consists of three $\pi$-rotations $\mathsf{C}_{2x}$, $\mathsf{C}_{2y}$, and $\mathsf{C}_{2z}$ in spin space. The first maps the spin components $(x,y,z)\mapsto(x,-y,-z)$, etc.

\end{enumerate}


\begin{table}[!t]
\caption{Character table of the double covers $\widetilde{\mathsf{C}}_{6\text{v}}$ and $\widetilde{\mathsf{D}}_{2}$ of the point groups $\mathsf{C}_{6\text{v}}$ and $\mathsf{C}_{6\text{v}}$, respectively. The thick horizontal line separates the regular from the spinor IRs.}\label{tab:Characters}
\begin{tabular}{l"ccccccccc}
$\widetilde{\mathsf{C}}_{6\text{v}}$&$\mathsf{E}$ & $\widetilde{\mathsf{E}}$ & $\{\mathsf{C}_2$,$\widetilde{\mathsf{C}}_2\}$ & $2\mathsf{C}_3$ & $2\widetilde{\mathsf{C}}_3$ & $2\mathsf{C}_6$ &  $2\widetilde{\mathsf{C}}_6$ & $\{3\sigma_{\text{d}}, 3\widetilde{\sigma}_{\text{d}}\}$ & $\{3\sigma_{\text{v}}, 3\widetilde{\sigma}_{\text{v}}\}$ \\
\thickhline
A$_1$ &1&1&1&1&1&1&1&1&1\\
A$_2$ &1&1&1&1&1&1&1&-1&-1\\
B$_1$ &1&1&-1&1&1&-1&-1&1&-1\\
B$_2$ &1&1&-1&1&1&-1&-1&-1&1\\
E$_1$ &2&2&-2&-1&-1&1&1&0&0\\
E$_2$ &2&2&2&-1&-1&-1&-1&0&0\\
\thickhline
E$_{1/2}$ &2&-2&0&1&-1&$\sqrt{3}$&$-\sqrt{3}$&0&0\\
E$_{3/2}$ &2&-2&0&1&-1&$-\sqrt{3}$&$\sqrt{3}$&0&0\\
E$_{5/2}$ &2&-2&0&-2&2&0&0&0&0\\
\thickhline
\end{tabular}
~~~~~~~~~~~~~~~~
\begin{tabular}{l"ccccc}
$\widetilde{\mathsf{D}}_2$&$\mathsf{E}$ &$\widetilde{\mathsf{E}}$ & $\{\mathsf{C}_{2z},\widetilde{\mathsf{C}}_{2z}\}$ & $\{\mathsf{C}_{2y},\widetilde{\mathsf{C}}_{2y}\}$ & $\{\mathsf{C}_{2x},\widetilde{\mathsf{C}}_{2x}\}$ \\
\thickhline
A &1&1&1&1&1\\
B$_1$ &1&1&1&-1&-1\\
B$_2$ &1&1&-1&1&-1\\
B$_3$ &1&1&-1&-1&1\\
\thickhline
E$_{1/2}$ &2&-2&0&0&0\\
\thickhline
\end{tabular}
\end{table}


\subsection{Finite clusters}
In our ED study we considered two clusters with periodic boundary conditions, one with 24 and another with 32 sites, with spanning vectors $(2\vec{t}_1\!-\!4\vec{t}_2,4\vec{t}_1\!-\!2\vec{t}_2)$ and $(2\vec{t}_1\!-\!4\vec{t}_2,4\vec{t}_1)$, respectively. These clusters are shown in Fig.~\ref{SMfig:EDclusters}~(a, c). The 24-site cluster has the full point group symmetry of the infinite lattice, i.e. $\widetilde{\mathsf{C}}_{6\text{v}}\times\widetilde{\mathsf{D}}_2$, whereas the 32-site cluster has the lower symmetry $\widetilde{\mathsf{C}}_{2\text{v}}\times\widetilde{\mathsf{D}}_2$, where $\widetilde{\mathsf{C}}_{2\text{v}}$ contains the reflection planes $(110)$ and $(1\bar{1}0)$. 
Turning to translational symmetry, the allowed momenta for each cluster are shown in Fig.~\ref{SMfig:EDclusters}(b, d). Both clusters accommodate the three $\vec{M}$ points of the Brillouin zone (BZ) and are therefore commensurate with all magnetic states of the phase diagram. The difference between the two clusters is that the three $\vec{M}$ points are degenerate for $N\!=\!24$ but not for $N\!=\!32$.

In our ED study we have exploited: i) translations, ii) the $\mathsf{C}_2$ subgroup of full $\mathsf{C}_{6\text{v}}$ point group (which is equivalent to the inversion $I$ in real space through the hexagon centers), and iii) the global spin inversion which maps the local $S_z$ basis states $|\!\uparrow\rangle\mapsto|\!\downarrow\rangle$. This operation is described by $\prod_i \sigma_i^x$, which is nothing else than the global $\pi$-rotation $\mathsf{C}_{2x}$ in spin space, divided by a phase factor $i^N$. Consequently, the energy eigenstates are labeled by: i) the momentum $\vec{k}$, ii) the parity under $\mathsf{C}_2$ (`e' for even, `o' for odd), and iii) the parity under $S_z$ spin inversion (`Sze' for even, `Szo' for odd).

\subsection{Symmetry spectroscopy of classical phases}\label{Sec:Spectr}
Here we derive the symmetry decomposition of the twelve magnetic states of region I and II of the phase diagram. As explained in the main paper, the other two regions, III and IV, map to I and II, respectively, by the  {\it hidden} duality of $H_{yxz}$ followed by a simultaneous change of sign in $K_1$ and $K_2$.

\subsubsection{Phase I}
In the following, $|\text{str},\alpha^{\bs{\beta}}\rangle$ denotes the stripy state with FM ladders running along the direction of the $\alpha$-bonds, and the spins pointing along $\bs{\beta}$ in spin space. The twelve magnetic states of region I of the phase diagram can be split into four groups: 
\small
\bea
&&
\mc{S}_1=\{|\text{str},x^{\vec{z}}\rangle, |\text{str},y^{\vec{x}}\rangle,|\text{str},z^{\vec{y}}\rangle\},~~
\overline{\mc{S}}_1=\{|\text{str},x^{-\vec{z}}\rangle, |\text{str},y^{-\vec{x}}\rangle,|\text{str},z^{-\vec{y}}\rangle\},\nonumber\\
&&
\mc{S}_2=\{|\text{str},y^{\vec{z}}\rangle, |\text{str},z^{\vec{x}}\rangle,|\text{str},x^{\vec{y}}\rangle\},~~
\overline{\mc{S}}_2=\{|\text{str},y^{-\vec{z}}\rangle, |\text{str},z^{-\vec{x}}\rangle,|\text{str},x^{-\vec{y}}\rangle\},\nonumber
\eea
\normalsize
Table~\ref{tab:Transf} shows how these twelve states transform under some of the symmetry operations of the group.  Let us first examine the translation group. We have, $\forall \bs{\beta}$:
\small 
\bea
&&
\mc{T}_{\vec{t}_1} \cdot |\text{str},x^{\bs{\beta}}\rangle =  |\text{str},x^{-\bs{\beta}}\rangle,
~~\mc{T}_{\vec{t}_2} \cdot |\text{str},x^{\bs{\beta}}\rangle = |\text{str},x^{-\bs{\beta}}\rangle, \nonumber\\
&&
\mc{T}_{\vec{t}_1} \cdot |\text{str},y^{\bs{\beta}}\rangle =  |\text{str},y^{-\bs{\beta}}\rangle,
~~\mc{T}_{\vec{t}_2} \cdot |\text{str},y^{\bs{\beta}}\rangle = |\text{str},y^{\bs{\beta}}\rangle, \nonumber\\
&&
\mc{T}_{\vec{t}_1} \cdot |\text{str},z^{\bs{\beta}}\rangle =  |\text{str},z^{\bs{\beta}}\rangle,
~~\mc{T}_{\vec{t}_2} \cdot |\text{str},z^{\bs{\beta}}\rangle = |\text{str},z^{-\bs{\beta}}\rangle.\nonumber
\eea
\normalsize 
Thus $\frac{1}{\sqrt{2}} \left( |\text{str},x^{\bs{\beta}}\rangle\!+\!|\text{str},x^{-\bs{\beta}}\rangle \right)$ transforms as $\vec{k}\!=\!0$ ($\bs{\Gamma}$ point) and 
$\frac{1}{\sqrt{2}} \left( |\text{str},x^{\bs{\beta}}\rangle\!-\!|\text{str},x^{-\bs{\beta}}\rangle \right)$ transforms as $\vec{k}\!=\!\frac{1}{a}(-\frac{\pi}{\sqrt{3}},\pi)\!\equiv\!\vec{M}_x$. 
Similarly, $\frac{1}{\sqrt{2}} \left( |\text{str},y^{\bs{\beta}}\rangle\!+\!|\text{str},y^{-\bs{\beta}}\rangle \right)$ transforms as $\vec{k}\!=\!0$ and $\frac{1}{\sqrt{2}} \left( |\text{str},y^{\bs{\beta}}\rangle\!-\!|\text{str},y^{-\bs{\beta}}\rangle \right)$ transforms as $\vec{k}\!=\!\frac{1}{a}(\frac{\pi}{\sqrt{3}},\pi)\!\equiv\!\vec{M}_y$, 
$\frac{1}{\sqrt{2}} \left( |\text{str},z^{\bs{\beta}}\rangle\!+\!|\text{str},z^{-\bs{\beta}}\rangle \right)$ transforms as $\vec{k}\!=\!0$, and $\frac{1}{\sqrt{2}} \left( |\text{str},z^{\bs{\beta}}\rangle\!-\!|\text{str},z^{-\bs{\beta}}\rangle \right)$ transforms as $\vec{k}\!=\!\frac{1}{a}(\frac{2\pi}{\sqrt{3}},0)\!\equiv\!\vec{M}_z$. Altogether:
\small
\bea
&&
\{|\text{str},x^{\vec{z}}\rangle, ~|\text{str},x^{-\vec{z}}\rangle \} \to  \bs{\Gamma} \oplus \vec{M}_x,~~~~~
\{|\text{str},y^{\vec{x}}\rangle, ~|\text{str},y^{-\vec{x}}\rangle \} \to  \bs{\Gamma} \oplus \vec{M}_y,~~~~~
\{|\text{str},z^{\vec{y}}\rangle, ~|\text{str},z^{-\vec{y}}\rangle \} \to  \bs{\Gamma} \oplus \vec{M}_z,\nonumber\\
&&
\{|\text{str},x^{-\vec{y}}\rangle, ~|\text{str},x^{\vec{y}}\rangle \} \to  \bs{\Gamma} \oplus \vec{M}_x,~~~~~
\{|\text{str},y^{-\vec{z}}\rangle, ~|\text{str},y^{-\vec{z}}\rangle \} \to  \bs{\Gamma} \oplus \vec{M}_y,~~~~~
\{|\text{str},z^{-\vec{x}}\rangle, ~|\text{str},z^{\vec{x}}\rangle \} \to  \bs{\Gamma} \oplus \vec{M}_z.\nonumber
\eea
\normalsize

\begin{table}[!t]
\caption{Transformations of the twelve states of region I under symmetry operations of the model. The phases that appear for some operations follow from the action of these operations on the single spin-1/2 wavefunctions $|\vec{n}\rangle$ corresponding to the spin pointing along $\vec{n}$ or $-\vec{n}\!\equiv\!\bar{\vec{n}}$:\\
\small
$\mathsf{C}_{6\text{v}}\cdot \{|\vec{x}\rangle,|\vec{y}\rangle,|\vec{z}\rangle,|\bar{\vec{x}}\rangle,|\bar{\vec{y}}\rangle,|\bar{\vec{z}}\rangle\} \mapsto \{-i |\vec{y}\rangle,e^{-i\pi/4}|\vec{z}\rangle,e^{-i\pi/4}|\vec{x}\rangle,|\bar{\vec{y}}\rangle,e^{-i\pi/4}|\bar{\vec{z}}\rangle,-e^{i\pi/4}|\bar{\vec{x}}\rangle\}$, \\
$\mathsf{C}_{2x}\cdot \{|\vec{x}\rangle,|\vec{y}\rangle,|\vec{z}\rangle,|\bar{\vec{x}}\rangle,|\bar{\vec{y}}\rangle,|\bar{\vec{z}}\rangle\} \mapsto \{-i |\vec{x}\rangle,|\bar{\vec{y}}\rangle,-i|\bar{\vec{z}}\rangle,i|\bar{\vec{x}}\rangle,-|\vec{y}\rangle,-i|\vec{z}\rangle\}$,\\
$(1\bar{1}0)\cdot \{|\vec{x}\rangle,|\vec{y}\rangle,|\vec{z}\rangle,|\bar{\vec{x}}\rangle,|\bar{\vec{y}}\rangle,|\bar{\vec{z}}\rangle\} \mapsto \{e^{-i\pi/4} |\bar{\vec{y}}\rangle,e^{i\pi/4}|\bar{\vec{x}}\rangle,-e^{i\pi/4}|\bar{\vec{z}}\rangle,-e^{-i\pi/4}|\vec{y}\rangle,-e^{i\pi/4}|\vec{x}\rangle,e^{-i\pi/4}|\vec{z}\rangle\}.$\\
For $N=24$ and $32$, the product of all these phase factors give $+1$.}\label{tab:Transf}
\begin{tabular}{l" l l l r" r r}
& $\mc{T}_{\vec{t}_1}$ (real sp.) & $\mc{T}_{\vec{t}_2}$ (real sp.) & $\mathsf{C}_2$ (real sp.) & $\mathsf{C}_{2x}$ (spin sp.) & $\mathsf{C}_6$ & $(1\bar{1}0)$ \\
\thickhline
$|\text{str},x^{\vec{z}}\rangle$ &$|\text{str},x^{-\vec{z}}\rangle$&$|\text{str},x^{-\vec{z}}\rangle$&$|\text{str},x^{\vec{z}}\rangle$&$(-1)^{N/2}|\text{str},x^{-\vec{z}}\rangle$&$(-1)^{N/2}|\text{str},y^{\vec{x}}\rangle$&$(-1)^{N/2}|\text{str},y^{-\vec{z}}\rangle$\\ 
$|\text{str},y^{\vec{x}}\rangle$ &$|\text{str},y^{-\vec{x}}\rangle$&$|\text{str},y^{\vec{x}}\rangle$&$|\text{str},y^{\vec{x}}\rangle$&$|\text{str},y^{\vec{x}}\rangle$&$(-i)^{N/2}|\text{str},z^{\vec{y}}\rangle$&$(-1)^{N/2}|\text{str},x^{-\vec{y}}\rangle$\\
$|\text{str},z^{\vec{y}}\rangle$ &$|\text{str},z^{\vec{y}}\rangle$ &$|\text{str},z^{-\vec{y}}\rangle$ &$|\text{str},z^{\vec{y}}\rangle$ &$(-1)^{N/2}|\text{str},z^{-\vec{y}}\rangle$ &$(-1)^{N/2}|\text{str},x^{\vec{z}}\rangle$ &$(-i)^{N/2}|\text{str},z^{-\vec{x}}\rangle$\\
&&&&&&\\
$|\text{str},x^{-\vec{z}}\rangle$ &$|\text{str},x^{\vec{z}}\rangle$&$|\text{str},x^{\vec{z}}\rangle$&$|\text{str},x^{-\vec{z}}\rangle$&$(-1)^{N/2}|\text{str},x^{\vec{z}}\rangle$&$(-1)^{N/2}|\text{str},y^{-\vec{x}}\rangle$&$(-1)^{N/2}|\text{str},y^{\vec{z}}\rangle$\\
$|\text{str},y^{-\vec{x}}\rangle$ &$|\text{str},y^{\vec{x}}\rangle$&$|\text{str},y^{-\vec{x}}\rangle$&$|\text{str},y^{-\vec{x}}\rangle$&$|\text{str},y^{-\vec{x}}\rangle$&$(-i)^{N/2}|\text{str},z^{-\vec{y}}\rangle$&$(-1)^{N/2}|\text{str},x^{\vec{y}}\rangle$\\
$|\text{str},z^{-\vec{y}}\rangle$ &$|\text{str},z^{-\vec{y}}\rangle$&$|\text{str},z^{\vec{y}}\rangle$&$|\text{str},z^{-\vec{y}}\rangle$&$(-1)^{N/2}|\text{str},z^{\vec{y}}\rangle$&$(-1)^{N/2}|\text{str},x^{-\vec{z}}\rangle$&$(-i)^{N/2}|\text{str},z^{\vec{x}}\rangle$\\
&&&&&&\\
$|\text{str},x^{\vec{y}}\rangle$ &$|\text{str},x^{-\vec{y}}\rangle$&$|\text{str},x^{-\vec{y}}\rangle$&$|\text{str},x^{\vec{y}}\rangle$&$(-1)^{N/2}|\text{str},x^{-\vec{y}}\rangle$&$(-1)^{N/2}|\text{str},y^{\vec{z}}\rangle$&$(-i)^{N/2}|\text{str},y^{-\vec{x}}\rangle$\\
$|\text{str},y^{\vec{z}}\rangle$ &$|\text{str},y^{-\vec{z}}\rangle$&$|\text{str},y^{\vec{z}}\rangle$&$|\text{str},y^{\vec{z}}\rangle$&$(-1)^{N/2}|\text{str},y^{-\vec{z}}\rangle$&$(-1)^{N/2}|\text{str},z^{\vec{x}}\rangle$&$(-1)^{N/2}|\text{str},x^{-\vec{z}}\rangle$\\
$|\text{str},z^{\vec{x}}\rangle$ &$|\text{str},z^{\vec{x}}\rangle$ &$|\text{str},z^{-\vec{x}}\rangle$ &$|\text{str},z^{\vec{x}}\rangle$ &$|\text{str},z^{\vec{x}}\rangle$ &$(-i)^{N/2}|\text{str},x^{\vec{y}}\rangle$ &$(-1)^{N/2}|\text{str},z^{-\vec{y}}\rangle$\\
&&&&&&\\
$|\text{str},x^{-\vec{y}}\rangle$ &$|\text{str},x^{\vec{y}}\rangle$&$|\text{str},x^{\vec{y}}\rangle$&$|\text{str},x^{-\vec{y}}\rangle$&$(-1)^{N/2}|\text{str},x^{\vec{y}}\rangle$&$|\text{str},y^{-\vec{z}}\rangle$&$(-i)^{N/2}|\text{str},y^{\vec{x}}\rangle$\\
$|\text{str},y^{-\vec{z}}\rangle$ &$|\text{str},y^{\vec{z}}\rangle$&$|\text{str},y^{-\vec{z}}\rangle$&$|\text{str},y^{-\vec{z}}\rangle$&$(-1)^{N/2}|\text{str},y^{\vec{z}}\rangle$&$(-1)^{N/2}|\text{str},z^{-\vec{x}}\rangle$&$(-1)^{N/2}|\text{str},x^{\vec{z}}\rangle$\\
$|\text{str},z^{-\vec{x}}\rangle$ &$|\text{str},z^{-\vec{x}}\rangle$&$|\text{str},z^{\vec{x}}\rangle$&$|\text{str},z^{-\vec{x}}\rangle$&$|\text{str},z^{-\vec{x}}\rangle$&$(-i)^{N/2}|\text{str},x^{-\vec{y}}\rangle$&$(-1)^{N/2}|\text{str},z^{\vec{y}}\rangle$\\
\thickhline
\end{tabular}
\end{table}

Next, let us examine the parities with respect to the $\mathsf{C}_2$ rotation in real space and the $\mathsf{C}_{2x}$ rotation in spin space. It is easy to see that the first symmetry is not broken by any of the twelve states, while the second is broken when $\bs{\beta}=\vec{y}$ and $\vec{z}$. So all twelve states are even with respect to $\mathsf{C}_2$, the $\bs{\beta}=\vec{x}$ are even with respect to $\mathsf{C}_{2x}$, while $\bs{\beta}=\vec{y}$ and $\vec{z}$ must decompose into both even and odd parities with respect to $\mathsf{C}_{2x}$. Altogether: 
\small
\bea\label{eq:Ia}
&&
\{|\text{str},x^{\vec{z}}\rangle, |\text{str},x^{-\vec{z}}\rangle \} \!\to\!  \bs{\Gamma}.e.\text{Sz}e \oplus \vec{M}_x.e.\text{Sz}o,~~~
\{|\text{str},x^{-\vec{y}}\rangle, |\text{str},x^{\vec{y}}\rangle \} \!\to\!  \bs{\Gamma}.e.\text{Sz}e \oplus \vec{M}_x.e.\text{Sz}o,\nonumber\\
&&
\{|\text{str},y^{\vec{x}}\rangle, |\text{str},y^{-\vec{x}}\rangle \} \!\to\!  \bs{\Gamma}.e.\text{Sz}e \oplus \vec{M}_y.e.\text{Sz}e,~~~
\{|\text{str},y^{-\vec{z}}\rangle, |\text{str},y^{\vec{z}}\rangle\}  \!\to\!  \bs{\Gamma}.e.\text{Sz}e \oplus \vec{M}_y.e.\text{Sz}o,\\
&&
\{|\text{str},z^{\vec{y}}\rangle, |\text{str},z^{-\vec{y}}\rangle \} \!\to\!  \bs{\Gamma}.e.\text{Sz}e \oplus \vec{M}_z.e.\text{Sz}o,~~~
\{|\text{str},z^{-\vec{x}}\rangle, |\text{str},z^{\vec{x}}\rangle \} \!\to\!  \bs{\Gamma}.e.\text{Sz}e \oplus \vec{M}_z.e.\text{Sz}e ~.\nonumber
\eea
\normalsize

\noindent{\bf `Extra' degeneracy at the $\vec{M}$ points for $N=24$.} 
The above quantum numbers for the $\vec{M}$ points are fully consistent with what we find in the low-energy spectra of Fig.~3~(a) of the main paper. For the symmetric, $N\!=\!24$ cluster, the three $\vec{M}$ points are degenerate due to the six-fold symmetry. However we see that the two sets of $\vec{M}$ points are also degenerate with respect to each other, i.e. we have a six-fold degeneracy. This extra degeneracy comes from the $\widetilde{\mathsf{D}}_{2}$ symmetry in spin space. To see this, let us relabel the spin inversion part of (\ref{eq:Ia}) using the actual IR of the group $\widetilde{\mathsf{D}}_2$ (see Table~\ref{tab:Characters}, right), instead of the parity with respect to $\mathsf{C}_{2x}$ (which contains less information about the state): 
\small
\bea\label{eq:Ib}
&&
\{|\text{str},x^{\vec{z}}\rangle, |\text{str},x^{-\vec{z}}\rangle \} \!\to\!  \bs{\Gamma}.e.\text{A} \oplus \vec{M}_x.e.\text{B}_1,~~~
\{|\text{str},x^{-\vec{y}}\rangle, |\text{str},x^{\vec{y}}\rangle \} \!\to\!  \bs{\Gamma}.e.\text{A} \oplus \vec{M}_x.e.\text{B}_2,\nonumber\\
&&
\{|\text{str},y^{\vec{x}}\rangle, |\text{str},y^{-\vec{x}}\rangle \} \!\to\!  \bs{\Gamma}.e.\text{A} \oplus \vec{M}_y.e.\text{B}_3,~~~
\{|\text{str},y^{-\vec{z}}\rangle, |\text{str},y^{\vec{z}}\rangle\}  \!\to\!  \bs{\Gamma}.e.\text{A} \oplus \vec{M}_y.e.\text{B}_1,\\
&&
\{|\text{str},z^{\vec{y}}\rangle, |\text{str},z^{-\vec{y}}\rangle \} \!\to\!  \bs{\Gamma}.e.\text{A} \oplus \vec{M}_z.e.\text{B}_2,~~~
\{|\text{str},z^{-\vec{x}}\rangle, |\text{str},z^{\vec{x}}\rangle \} \!\to\!  \bs{\Gamma}.e.\text{A} \oplus \vec{M}_z.e.\text{B}_3 ~.\nonumber
\eea
\normalsize
We see that the two states belonging to a given $\vec{M}$ point transform differently under $\widetilde{\mathsf{D}}_2$, so the Hamiltonian does not couple the two states. Yet, these states are mapped to each other by one of the reflection planes of $\widetilde{\mathsf{C}}_{6\text{v}}$, so they must be degenerate, leading to an overall six-fold degeneracy at the $\vec{M}$ points. 
\\

\noindent{\bf Degeneracies at the $\bs{\Gamma}$ point for $N=24$.} The little group of the $\bs{\Gamma}$ point is the full point group $\widetilde{\mathsf{C}}_{6\text{v}}\times\widetilde{\mathsf{D}}_{2}$. However, all of the above six states that belong to the $\bs{\Gamma}$ point belong to the identity IR of $\widetilde{\mathsf{D}}_2$, so it is enough to decompose them with respect to the $\widetilde{\mathsf{C}}_{6\text{v}}$ part of the little group. To this end we use the well known formula from group theory~\cite{TinkhamSM}
\be
m_\alpha = \frac{1}{|\widetilde{\mathsf{C}}_{6\text{v}}|} \sum_{{\bf g}\in \widetilde{\mathsf{C}}_{6\text{v}}} \chi^\alpha({\bf g}) X({\bf g})^\ast~,
\ee
which gives the number of times $m_\alpha$ that the $\alpha$-th IR of $\widetilde{\mathsf{C}}_{6\text{v}}$ appears in the decomposition of the $6\times6$ representation formed by the six states belonging to the $\bs{\Gamma}$ point. Here $X({\bf g})$ gives the character of this representation, while $\chi^\alpha({\bf g})$ is the character of the $\alpha$-th IR of $\widetilde{\mathsf{C}}_{6\text{v}}$, see Table~\ref{tab:Characters}~(left). From Table~\ref{tab:Transf} it follows that $X({\bf g})$ is finite only for the elements $E$, $\widetilde{E}$, $\mathsf{C}_2$, and $\widetilde{\mathsf{C}}_2$, and using the characters of Table~\ref{tab:Characters}~(left) we find that the only finite $m_\alpha$ are the following: $m_{A_1}=m_{A_2}=1$, $m_{E_2}=2$, namely
\be\label{eq:IG}
6\bs{\Gamma}\to \text{A}_1\oplus \text{A}_2\oplus 2\text{E}_2~.
\ee
i.e. we expect two singlets and two doublets. All states are found in the low-energy spectra shown in Fig.~3~(a) of the main paper, where the degeneracy of the E$_2$ levels has been confirmed numerically.

\subsubsection{Phase II}
Here we denote by $|\text{zig},\alpha\alpha'^{\bs{\beta}}\rangle$ the zigzag state with FM lines formed by consecutive $\alpha$ and $\alpha'$ type of bonds, and the spins pointing along $\bs{\beta}$ in spin space. The twelve magnetic states of region II can be split into four groups: 
\small
\bea
&&
\mc{S}_3=\{|\text{zig},yz^{\vec{z}}\rangle, |\text{zig},zx^{\vec{x}}\rangle,|\text{zig},xy^{\vec{y}}\rangle\},~~
\overline{\mc{S}}_3=\{|\text{zig},yz^{-\vec{z}}\rangle, |\text{zig},zx^{-\vec{x}}\rangle,|\text{zig},xy^{-\vec{y}}\rangle\},\nonumber\\
&&
\mc{S}_4=\{|\text{zig},zx^{\vec{z}}\rangle, |\text{zig},xy^{\vec{x}}\rangle,|\text{zig},yz^{\vec{y}}\rangle\},~~
\overline{\mc{S}}_4=\{|\text{zig},zx^{-\vec{z}}\rangle, |\text{zig},xy^{-\vec{x}}\rangle,|\text{zig},yz^{-\vec{y}}\rangle\},\nonumber
\eea
\normalsize
Under $\mathsf{T}$ and $\mathsf{C}_{2x}$ in spin space, these states transform in analogous way with the twelve states of region I, see (\ref{eq:Ia}). The difference is that the present states break the $\mathsf{C}_2$ rotation around the hexagon centers, and therefore the decomposition will contain both even and odd parities with respect to $\mathsf{C}_2$. Specifically,
\small
\bea\label{eq:II}
&&
\{|\text{zig},yz^{\vec{z}}\rangle, |\text{zig},yz^{-\vec{z}}\rangle \} \!\to\!  \bs{\Gamma}.e.\text{Sz}e \oplus \vec{M}_x.o.\text{Sz}o,~~~
\{|\text{zig},yz^{-\vec{y}}\rangle, |\text{zig},yz^{\vec{y}}\rangle \} \!\to\!  \bs{\Gamma}.e.\text{Sz}e \oplus \vec{M}_x.o.\text{Sz}o,\nonumber\\
&&
\{|\text{zig},zx^{\vec{x}}\rangle, |\text{zig},zx^{-\vec{x}}\rangle \} \!\to\!  \bs{\Gamma}.e.\text{Sz}e \oplus \vec{M}_y.o.\text{Sz}e,~~~
\{|\text{zig},zx^{-\vec{z}}\rangle, |\text{zig},zx^{\vec{z}}\rangle\}  \!\to\!  \bs{\Gamma}.e.\text{Sz}e \oplus \vec{M}_y.o.\text{Sz}o,\\
&&
\{|\text{zig},xy^{\vec{y}}\rangle, |\text{zig},xy^{-\vec{y}}\rangle \} \!\to\!  \bs{\Gamma}.e.\text{Sz}e \oplus \vec{M}_z.o.\text{Sz}o,~~~
\{|\text{zig},xy^{-\vec{x}}\rangle, |\text{zig},xy^{\vec{x}}\rangle \} \!\to\!  \bs{\Gamma}.e.\text{Sz}e \oplus \vec{M}_z.o.\text{Sz}e ~.\nonumber
\eea
\normalsize
In analogy with region I, for the symmetric 24-site cluster, the six states belonging to the $\vec{M}$ points are degenerate due to the additional $\widetilde{\mathsf{D}}_2$ symmetry, and the six states belonging to the $\bs{\Gamma}$ point decompose as in (\ref{eq:IG}), namely $6\bs{\Gamma}\to \text{A}_1\oplus \text{A}_2\oplus 2\text{E}_2$. Again, all states are found in the low-energy spectra shown in Fig.~3~(a) of the main paper.

\subsubsection{Special points $\psi=\pm\pi/2$: Different ground state structure for $N\!=\!24$ and $N\!=\!32$}

As shown in Figs.~3(a) and (b) of the main text, the ED results are broadly independent of system size but significant differences between the two cluster sizes are apparent for the GS structure near $\psi\!=\!\pm\pi/2$. 
The reason behind this difference lies in the different point group symmetry of the two clusters. The 24-site cluster has the full point group symmetry of the infinite lattice, whereas the 32-site cluster does not. This is also true for the two triangular sublattices of each cluster at $\psi=\pm\pi/2$, where they become independent from each other.

Due to the high symmetry, each of the 12-site sublattices of the 24-site cluster have a two-fold degenerate ground state at $\psi=\pm\pi/2$; let us denote them by $|\alpha\rangle$ and $|\beta\rangle$. On the other hand, the lower symmetry of the 16-site sublattices of the 32-site cluster leads to a single, non-degenerate ground state; let us denote it by $|\gamma\rangle$. 
Now, the global ground state structure of the two clusters at $\psi=\pm\pi/2$ follows simply by taking the tensor product of the ground state manifolds in each sublattice. The 24-site cluster has four ground states:
\begin{equation}
|\alpha\rangle_{\text{sub1}}\otimes|\alpha\rangle_{\text{sub2}},~~
|\beta\rangle_{\text{sub1}}\otimes|\beta\rangle_{\text{sub2}},~~
|\alpha\rangle_{\text{sub1}}\otimes|\beta\rangle_{\text{sub2}},~~
|\beta\rangle_{\text{sub1}}\otimes|\alpha\rangle_{\text{sub2}}.
\end{equation}
The first two states belong to the representation $\Gamma.e.\text{Sz}e$, i.e. they have even parity with respect to inversion through the middle of the hexagons (this operation maps one sublattice to the other), and the same is true for the combination $\frac{1}{\sqrt{2}}(|\alpha\rangle_{\text{sub1}}\otimes|\beta\rangle_{\text{sub2}}+|\beta\rangle_{\text{sub1}}\otimes|\alpha\rangle_{\text{sub2}})$. The remaining, antisymmetric combination, $\frac{1}{\sqrt{2}}(|\alpha\rangle_{\text{sub1}}\otimes|\beta\rangle_{\text{sub2}}-|\beta\rangle_{\text{sub1}}\otimes|\alpha\rangle_{\text{sub2}})$, belongs to $\Gamma.o.\text{Sz}e$, i.e. it has odd parity. This is in perfect agreement with the ED data.

For the 32-site cluster on the other hand, there is only one global ground state, namely $|\gamma\rangle_{\text{sub1}}\otimes|\gamma\rangle_{\text{sub2}}$, which has even parity, again in agreement with the ED data.

Of course, as we discuss in the main text, in the thermodynamic limit a large number of states ($12^2$) will collapse to the ground state, which is how the corresponding symmetry-broken (classical) states are eventually formed.

\subsection{`Symmetrized' spin structure factor and spin length}
Here we discuss the `symmetrized' spin structure factor $\widetilde{S}(\vec{Q})$ and explain the overall normalization factor that we use to extract the spin length. As we discuss in the main text, NN ladders do not couple by the symmetry $H_{xyz}$, and so the quantum ground state of a finite cluster contains both relative orientations of the two sets of ladders $L_1$ and $L_2$ with equal amplitude. As a result, the spin-spin correlations between two spins that belong to $L_1$ and $L_2$ are zero for any finite cluster. If we wish to calculate the local spin lengths from the ground state spin-spin correlation data we can calculate the `symmetrized' spin structure factor for one of the two subsets of ladders only, say $L_1$:
\be
\mc{S}_1(\vec{Q})=\frac{1}{N_1^2} \sum_\alpha \sum_{\vec{r},\vec{r}'\in L_1} \langle S^{\alpha}_{\vec{r}} S^{\alpha}_{\vec{r}'} \rangle e^{i\vec{Q}^{(\alpha)}\cdot(\vec{r}-\vec{r}')},
\ee
where $N_1=N/2$ is the number of sites inside the sublattice $L_1$, and $\vec{Q}^{(a)}$ is the ordering wavevector corresponding to the spin component $\alpha=\{x,y,z\}$.  
By translation symmetry, 
\be
\langle S^{\alpha}_{\vec{r}} S^{\alpha}_{\vec{r}'}\rangle=\langle S^{\alpha}_{\vec{r}+\bs{\delta}} S^{\alpha}_{\vec{r}'+\bs{\delta}}\rangle
\Rightarrow
\mc{S}_1(\vec{Q})=\frac{1}{N_1} \sum_{\alpha}\sum_{\vec{r}\in L_1} \langle S^{\alpha}_{0} S^{\alpha}_{\vec{r}} \rangle e^{i\vec{Q}^{(\alpha)}\cdot\vec{r}},
\ee
where we have chosen a reference site $\vec{r}'=0$. The local spin length $m$ is then given by $m^2 =\frac{2}{N}\mc{S}_1(\vec{Q})$.

By contrast, the corresponding `symmetrized' spin structure factor of the full lattice $\mc{S}(\vec{Q})$, defined by 
\be
\mc{S}(\vec{Q})=\frac{1}{N^2} \sum_{\alpha} \sum_{\vec{r},\vec{r}' ~\in ~L_1\cup L_2} \langle S^{\alpha}_{\vec{r}} S^{\alpha}_{\vec{r}'} \rangle e^{i\vec{Q}^{(\alpha)}\cdot(\vec{r}-\vec{r}')},
\ee
would give in the present case
\be
\mc{S}(\vec{Q})=\frac{1}{2}\mc{S}_1(\vec{Q}),
\ee
and the corresponding local spin lengths would be off by a factor of $\sqrt{2}$.

\section{Pseudofermion functional renormalization group (PFFRG) approach}\label{PFFRG}
In addition to ED we studied the $K_1$-$K_2$ honeycomb model using the pseudofermion functional renormalization group (PFFRG) approach. Rewriting the spin operators in terms of Abrikosov auxiliary fermions, the resulting fermionic model can be efficiently treated using a one loop functional renormalization group procedure. This technique calculates diagrammatic contributions to the spin-spin correlation function in infinite order in the exchange couplings, including terms in different interaction channels: The inclusion of direct particle-hole terms insures the correct treatment of the large spin limit $S \rightarrow \infty$ while the crossed particle-hole and particle-particle terms lead to exact results in the large $N$ limit. This allows to study the competition between magnetic order tendencies and quantum fluctuations in an unbiased way. For details we refer to reader to Ref.~[\onlinecite{PhysRevB.81.144410SM}].

The PFFRG method calculates the {\it static} spin-structure factor as given by
\begin{equation}
\chi^{\alpha\beta}(\mathbf{k})=\int_{0}^{\infty}d\tau \langle T_{\tau}\{S^{\alpha}(-\mathbf{k},\tau)S^{\beta}(\mathbf{k},0)\}\rangle\,,\label{spin_structure}
\end{equation} 
with
\begin{equation}
S^{\alpha}(\mathbf{k},\tau)=\frac{1}{\sqrt{N}}\sum_i e^{-i\mathbf{k}\mathbf{r}_i}e^{H\tau}S_i^ze^{-H\tau}\,,
\end{equation} 
where $\tau$ denotes the imaginary time and $T_\tau$ is the corresponding time-ordering operator. Being able to treat large system sizes (calculations for the $K_1$-$K_2$ model are performed for a spin cluster with 265 sites) the PFFRG yields results close to the thermodynamic limit. Fig.~\ref{fig:frg} shows three representative plots for the momentum resolved spin-structure factor $\chi^{zz}(\mathbf{k})$ in the Kitaev spin-liquid phase in the vicinity of $\psi=0$. While in the exact Kitaev limit $\psi=0$ the PFFRG reproduces the well-known nearest neighbor correlations as indicated by a single harmonics profile of the spin-structure factor, deviations from $\psi=0$ lead to longer-range correlations and a more diverse spin-structure factor.
\begin{figure*}[t]
\centering
\includegraphics[width=0.9\linewidth]{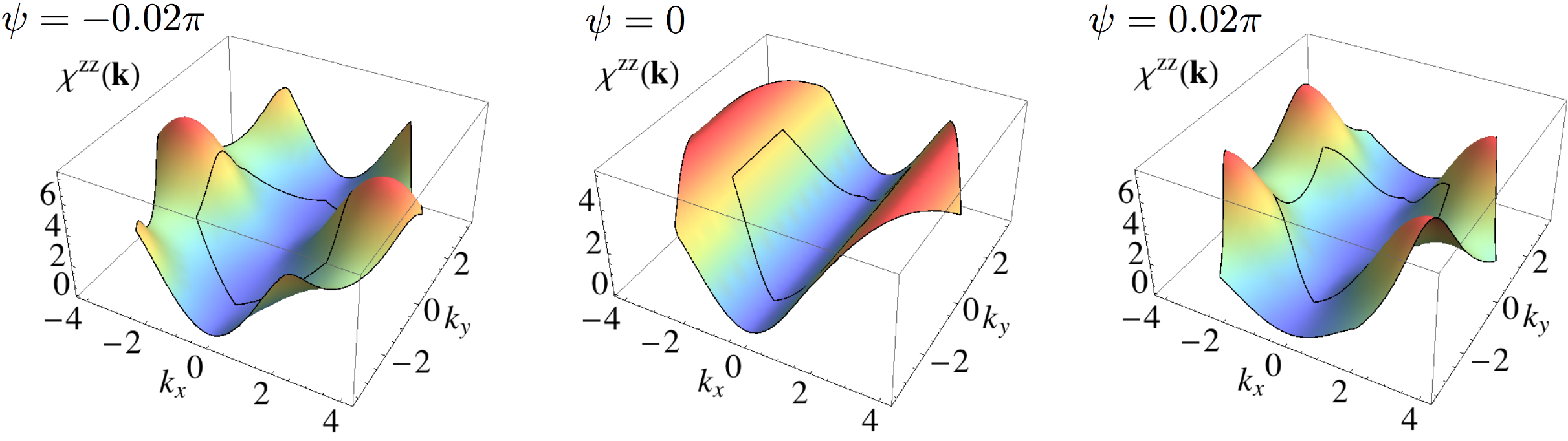}
\caption{Static spin-structure factor $\chi^{zz}(\mathbf{k})$ plotted in the extended Brillouin zone (black lines inside the plotted region mark the boundaries of first Brillouin zone) for various values of $\psi$ in the Kitaev spin-liquid phase. Note that $\chi^{xx}(\mathbf{k})$ ($\chi^{yy}(\mathbf{k})$) are related to $\chi^{zz}(\mathbf{k})$ by clockwise (counterclockwise) $2\pi/3$-rotations in k-space.}
\label{fig:frg}
\end{figure*}

\section{Strong-coupling expansion}\label{Sec:Heff}
Here we provide some technical details on the derivation of the effective model around the strong coupling limit of $K_{1}^{x(y)}\!=\!K_{2}^{x(y)}\!=\!0$. In this limit we have $n_{\text{lad}}$ decoupled vertical ladders (which are the ladders made of the $zz$-bonds), leading to a sub-extensive ground state degeneracy. The ordering pattern within each individual vertical ladder is fixed (up to a global sign) by the signs of $K_1^z$ and $K_2^z$. 
The GS degeneracy is lifted by the transverse perturbations $K_1^{x(y)}$ and $K_2^{x(y)}$, which give rise to effective couplings between the ladders (or more accurately between NNN ladders, as discussed in the main paper). These couplings can be found by degenerate perturbation theory. Let us denote by $H_0$ the sum of all $K_1^z$ interactions and by $V$ the sum of all remaining terms of the model. 
In the following we define the strong coupling parameter $r$ to be the ratio between $K_{1,2}^{x(y)}$ and $K_{1,2}^z$, as in the main text.

To write down the effective Hamiltonian we should define the corresponding Hilbert space on which it acts. Obviously, for $K_2^z\!\neq\!0$ this is the ground state manifold of $H_0$, namely the $2^{n_{\text{lad}}}$ states corresponding to all possible relative orientations of the vertical ladders. However, special care must be taken at $K_2^z\!=\!0$ where different rungs of a given vertical ladder do not interact with each other and the relevant Hilbert space is enlarged from $2^{n_{\text{lad}}}$ to $2^{N/2}$, where $N$ is the number of sites. To treat both $K_2^z\!\neq\!0$ and $K_2^z\!=\!0$ cases at once we must then take the enlarged manifold of $2^{N/2}$ states. 

With this definition of the target Hilbert space, the very first term of the effective Hamiltonian is a first-order coupling between the rungs which is proportional to $K_2^z$, which fixes (for $K_2^z\!\neq\!0$) the relative orientation of different rungs within each vertical ladder. It is easy to see that the remaining degeneracy between different ladders is lifted in fourth-order in $V$. The effective Hamiltonian (up to fourth order) is then described by the expression 
\small
\be\label{eq:Heff-1}
\mc{H}_{\text{eff}} = \mc{H}_{1} + P V R V R V R V P\,,
\ee
\normalsize
where $\mc{H}_1$ contains the $K_2^z$ terms, $P$ is the projection into the enlarged manifold of $2^{N/2}$ states discussed above, and $R\!=\!\frac{1-P}{E_0-H_0}$ is the resolvent, where $E_0\!=\!(-|K_1|/2-|K_2|)N$ is the ground state energy at $V\!=\!0$. By expanding the different terms of $V$ in (\ref{eq:Heff-1}) we get three types of loop-four virtual processes, that involve: i) only NN perturbations $K_1^{x(y)}$ (Sec.~\ref{Sec:i}), ii) only NNN perturbations $K_2^{x(y)}$ (Sec.~\ref{Sec:ii}), and iii) both $K_1^{x(y)}$ and $K_2^{x(y)}$ perturbations (Sec.~\ref{Sec:iii}). 

\begin{figure}[!t]
\includegraphics[width=0.5\linewidth,angle=90,clip=true,trim=120 0 100 0]{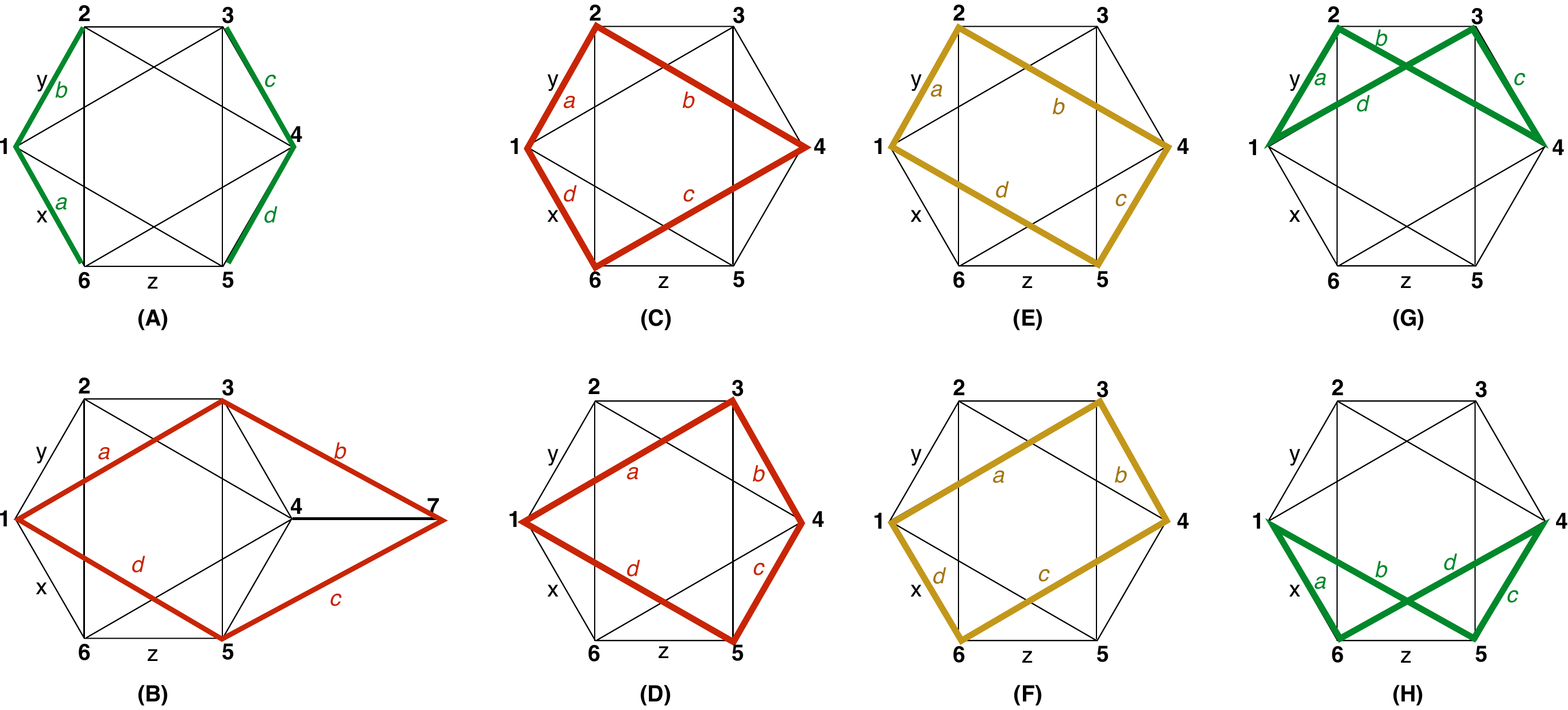}
\caption{\label{fig:EffJs} Effective couplings arising from only NN perturbations (A), only NNN perturbations (B), and both NN and NNN perturbations (C-H).} 
\end{figure}

\subsection{Effective terms arising from $K_1^{x(y)}$ only (Toric code terms)}\label{Sec:i}
The $K_1^{x(y)}$ perturbations give rise to intra-ladder, six-body terms of the form $J_W \hat{W}_p$, where $\hat{W}_p$ is Kitaev's~\cite{kitaev06SM} flux operator: 
\small
\be
\hat{W}_p=2^6 S_1^zS_2^yS_3^xS_4^zS_5^yS_6^x~,
\ee
\normalsize
where $1$-$6$ label clockwise the six sites of the hexagon plaquette $p$, as shown in Fig.~\ref{fig:EffJs}~(A). To find $J_W$ in fourth order in $r$, it suffices to consider one hexagon only. Let us denote the local configuration of this hexagon in any of the ground states at $r\!=\!0$ by $|\sigma_1,\sigma_2,\sigma_3,\sigma_4,\sigma_5,\sigma_6\rangle $, with the spin projections $S_j^z\!=\!\frac{1}{2}\sigma_j$, and $\sigma_1\sigma_6 \!=\! \sigma_3\sigma_4 \!=\!-\text{sgn}(K_1^z)$. In this case, the perturbation $V\!=\!A$ can be written as (see Fig.~\ref{fig:EffJs}):
\small
\be\label{eq:A}
A=A_a+A_b+A_c+A_d=K_1^x S_1^xS_6^x+K_1^y S_1^yS_2^y + K_1^x S_3^xS_4^x + K_1^y S_4^yS_5^y\,,
\ee
\normalsize
and Eq. (\ref{eq:Heff-1}) contains 24 terms in total, which have the form 
\small
\be
\mc{H}_{\text{eff}}^{(A,dcba)} = PA_dRA_cRA_bRA_aP, ~~~~\text{etc.}\nonumber
\ee
\normalsize
In the following we define $\mu\!=\!(K_1^x K_1^y)^2$ and use the relations $S^x |\sigma\rangle\!=\!\frac{1}{2}|\!-\sigma\rangle$ and  $S^y |\sigma\rangle\!=\!\frac{i \sigma}{2}|\!-\sigma\rangle$. The energy excitations of various intermediate states are 
\small
\bea
&&\Delta_{12}=\Delta_{16}=\Delta_{34}=\Delta_{45}=-|K_1^z|-2|K_2^z|, \nonumber\\
&&\Delta_{26}=\Delta_{35}=-|K_1^z|-|K_2^z|,\nonumber\\
&&\Delta_{1635}=\Delta_{1235}=\Delta_{2634}=\Delta_{2645}=-|K_1^z|-3|K_2^z|,\nonumber\\
&&\Delta_{1634}=\Delta_{1245}=-2|K_1^z|-4|K_2^z|,\nonumber\\
&&\Delta_{1234}=\Delta_{1645}=-|K_1^z|-4|K_2^z|,\nonumber
\eea
\normalsize
Let us first consider the terms of the type
\small
\bea
\mc{H}_{\text{eff}}^{(A,dcba)} |\sigma_1,\sigma_2,\sigma_3,\sigma_4,\sigma_5,\sigma_6\rangle 
\!=\!\frac{\mu~ \sigma_1\sigma_2\sigma_4\sigma_5}{4^4\Delta_{45}\Delta_{35}\Delta_{1235}} |\sigma_1,-\sigma_2,-\sigma_3,\sigma_4,-\sigma_5,-\sigma_6\rangle \,.\nonumber
\eea
\normalsize
The final state is not the same as the initial one, but belongs to the enlarged manifold of $2^{N/2}$ states, so this is a valid process. The operator that does the job is:
\small
\be
\frac{\mu}{4\Delta_{45}\Delta_{35}\Delta_{1235}}  S_1^z S_2^y S_3^x S_4^z S_5^y S_6^x \!=\! \frac{\mu}{2^8\Delta_{45}\Delta_{35}\Delta_{1235}}  \hat{W}_p\,.\nonumber
\ee
\normalsize
This result can be also found right away by taking 
\small
\be
\mc{H}_{\text{eff}}^{(A,dcba)}  \!=\! \mc{H}_{\text{eff}}^{(A,abcd)}  \!=\!
\mu P S_1^xS_6^x R S_1^yS_2^y R S_3^xS_4^x R S_4^yS_5^y P \to 
\frac{\mu}{D_1} \underbrace{(S_1^xS_1^y)}_{\frac{i}{2}S_1^z} S_2^y S_3^x \underbrace{(S_4^x S_4^y)}_{\frac{i}{2}S_4^z} S_5^y S_6^x  
\!=\!-\frac{\mu}{2^8D_{1}}  \hat{W}_p~,\nonumber
\ee
\normalsize
with $D_{1}\!=\!\Delta_{45}\Delta_{35}\Delta_{1235}$. 
Similarly
\small
\bea\nonumber
\mc{H}_{\text{eff}}^{(A,cdba)} \!=\! \mc{H}_{\text{eff}}^{(A,abdc)} \!=\! 
\mc{H}_{\text{eff}}^{(A,dcab)} \!=\!  \mc{H}_{\text{eff}}^{(A,bacd)} \!=\!
-\mc{H}_{\text{eff}}^{(A,cdab)} \!=\! - \mc{H}_{\text{eff}}^{(A,badc)} \!=\! \frac{\mu}{2^8D_{1}}  \hat{W}_p~,
\eea
\normalsize
where we used $\Delta_{34}\Delta_{35}\Delta_{1235}\!=\!\Delta_{45}\Delta_{35}\Delta_{1235}\!=\!\Delta_{45}\Delta_{35}\Delta_{1635}\!=\!\Delta_{45}\Delta_{35}\Delta_{1235}\!=\!\Delta_{34}\Delta_{35}\Delta_{1635}\!=\!\Delta_{45}\Delta_{35}\Delta_{1235}\!=\!D_{1}$. So the eight processes $\{abcd,bacd,abdc,badc\}$ and $\{dcba,dcab,cdba,cdab\}$ cancel each other out.

Next come the processes:
\small
\bea
\mc{H}_{\text{eff}}^{(A,acbd)} \!=\!  \mc{H}_{\text{eff}}^{(A,dbca)} \!=\! \mu P S_4^yS_5^y R S_1^yS_2^y R S_3^xS_4^x  R  S_1^xS_6^x  P 
\to \frac{\mu}{D_2} \underbrace{(S_1^yS_1^x)}_{-\frac{i}{2}S_1^z}  S_2^y  S_3^x \underbrace{(S_4^yS_4^x)}_{-\frac{i}{2}S_4^z} S_5^y S_6^x  
= -\frac{\mu}{2^8D_2} \hat{W}_p~,\nonumber
\eea
\normalsize
with $D_2\!=\!\Delta_{16}\Delta_{1634}\Delta_{2634}$. Similarly, 
\small
\bea
\mc{H}_{\text{eff}}^{(A,cabd)} \!=\!  \mc{H}_{\text{eff}}^{(A,dbac)}\!=\!
\mc{H}_{\text{eff}}^{(A,acdb)} \!=\! \mc{H}_{\text{eff}}^{(A,bdca)}\!=\! 
\mc{H}_{\text{eff}}^{(A,cadb)} \!=\!  \mc{H}_{\text{eff}}^{(A,bdac)} \!=\! -\frac{\mu}{2^8D_2}  \hat{W}_p,\nonumber
\eea
\normalsize
where we used $\Delta_{34}\Delta_{1634}\Delta_{2634}\!=\!\Delta_{16}\Delta_{1634}\Delta_{1635}\!=\!\Delta_{34}\Delta_{1634}\Delta_{1635}\!=\!D_2$. These eight processes $\{cabd,acbd,cadb,acdb\}$ and $\{dbac,dbca,bdac,bdca\}$ give the same contribution and, thus, do not cancel out. 

Finally, there are the processes
\small
\bea
\mc{H}_{\text{eff}}^{(A,cbad)} \!=\!  \mc{H}_{\text{eff}}^{(A,dabc)} 
\!=\!\mu P S_4^yS_5^y R  S_1^xS_6^x  R S_1^yS_2^y R S_3^xS_4^x   P 
\to \frac{\mu}{D_3} \underbrace{(S_1^xS_1^y)}_{\frac{i}{2}S_1^z}  S_2^y  S_3^x \underbrace{(S_4^yS_4^x)}_{-\frac{i}{2}S_4^z} S_5^y S_6^x  
\!=\! +\frac{\mu}{2^8D_3}  \hat{W}_p~,\nonumber
\eea
\normalsize
with $D_3\!=\!\Delta_{34}\Delta_{1234}\Delta_{2634}$. Similarly 
\small
\bea
\mc{H}_{\text{eff}}^{(A,bcad)} \!=\!  \mc{H}_{\text{eff}}^{(A,dacb)} \!=\! 
\mc{H}_{\text{eff}}^{(A,cbda)} \!=\!  \mc{H}_{\text{eff}}^{(A,adbc)} \!=\! 
\mc{H}_{\text{eff}}^{(A,bcda)} \!=\!  \mc{H}_{\text{eff}}^{(A,adcb)} \!=\! +\frac{\mu}{2^8D_3}  \hat{W}_p~,\nonumber
\eea
\normalsize
where we used $\Delta_{12}\Delta_{1234}\Delta_{2634}\!=\!\Delta_{34}\Delta_{1234}\Delta_{1235}\!=\!\Delta_{12}\Delta_{1234}\Delta_{1235}\!=\!D_3$. So the eight processes $\{cbad,bcad,cbda,bcda\}$ and $\{dabc,dacb,adbc,adcb\}$ also do not cancel out. 
Altogether:
\small
\be
\mc{H}_{\text{eff}}^{(A)} \!=\!8 \mc{H}_{\text{eff}}^{(A,acbd)} +8\mc{H}_{\text{eff}}^{(A,cbad)}\!=\!
\frac{\mu}{2^5} (\frac{1}{D_3}-\frac{1}{D_2}) \hat{W}_p\!=\!\frac{\mu(\Delta_{1634}-\Delta_{1234})}{2^5\Delta_{34}\Delta_{1634}\Delta_{1635}\Delta_{1234}} \hat{W}_p~.\nonumber 
\ee
\normalsize
We have $\Delta_{1634}-\Delta_{1234}\!=\!-|K_1^z|$, and therefore
\small
\be
\boxed{\mc{H}_{\text{eff}}^{(A)} = J_{W}\hat{W}_p}, ~~~
\boxed{J_W=\frac{-\mu |K_1^z|}{2^6(|K_1^z|+2|K_2^z|)^2(|K_1^z|+3|K_2^z|)(|K_1^z|+4|K_2^z|)}}~.
\ee
\normalsize
For $K_2^z\!=\!0$ we get $J_W\!=\!-\frac{(K_1^xK_1^y)^2}{2^6|K_1^z|^3}$, which agrees with the result obtained by Kitaev~\cite{kitaev06SM}.

\subsection{Effective terms arising from $K_2^{x(y)}$ only.}\label{Sec:ii}
Consider three consecutive ladders in the honeycomb lattice. We will show that the $K_2^{x(y)}$ terms give rise to an effective NNN inter-ladder coupling of the form $J_{1}S_1^z S_7^z$, see Fig.~\ref{fig:EffJs}~(B). In this case, the perturbation $V\!=\!B$ is given by (see Fig.~\ref{fig:EffJs}):
\small
\be\label{eq:B}
B=B_a+B_b+B_c+B_d=K_2^x S_1^xS_3^x+K_2^y S_3^yS_7^y+K_2^x S_5^xS_7^x+K_2^y S_1^yS_5^y~.
\ee   
\normalsize
Again, Eq.~(\ref{eq:Heff-1}) gives 24 relevant contributions. 
In the following we define $\lambda\!=\!(K_2^x K_2^y)^2$, and use the relation $\sigma_3\sigma_5\!=\!-\text{sgn}(K_2^z)$. We also introduce the excitation energies of various intermediate virtual states: 
\small
\be
\Delta_{13}\!=\!\Delta_{17}\!=\!\Delta_{15}\!=\!\Delta_{37}\!=\!\Delta_{57}\!=\!-|K_1^z|-2|K_2^z|, ~~ 
\Delta_{35}\!=\!-|K_1^z|-|K_2^z|, ~~
\Delta_{1357}\!=\!-2|K_1^z|-3|K_2^z|~.\nonumber
\ee
\normalsize
We find:
\small
\bea
\mc{H}_{\text{eff}}^{(B,abcd)} |\sigma_1,\sigma_2,\sigma_3,\sigma_4,\sigma_5,\sigma_6,\sigma_7\rangle &=& 
+\frac{\lambda\sigma_1\sigma_7}{4^4\Delta_{13}\Delta_{17}\Delta_{15}} \text{sgn}(K_2^z) |\sigma_1,\sigma_2,\sigma_3,\sigma_4,\sigma_5,\sigma_6,\sigma_7\rangle~,\nonumber\\
\mc{H}_{\text{eff}}^{(B,abdc)} |\sigma_1,\sigma_2,\sigma_3,\sigma_4,\sigma_5,\sigma_6,\sigma_7\rangle &=& 
-\frac{\lambda\sigma_7\sigma_1}{4^4\Delta_{13}\Delta_{17}\Delta_{57}}  \text{sgn}(K_2^z) |\sigma_1,\sigma_2,\sigma_3,\sigma_4,\sigma_5,\sigma_6,\sigma_7\rangle~,\nonumber\\
\mc{H}_{\text{eff}}^{(B,bacd)} |\sigma_1,\sigma_2,\sigma_3,\sigma_4,\sigma_5,\sigma_6,\sigma_7\rangle &=& 
= -\frac{\lambda\sigma_1\sigma_7}{4^4\Delta_{37}\Delta_{17}\Delta_{15}} \text{sgn}(K_2^z) |\sigma_1,\sigma_2,\sigma_3,\sigma_4,\sigma_5,\sigma_6,\sigma_7\rangle~,\nonumber\\
\mc{H}_{\text{eff}}^{(B,badc)} |\sigma_1,\sigma_2,\sigma_3,\sigma_4,\sigma_5,\sigma_6,\sigma_7\rangle &=& 
+\frac{\lambda\sigma_1\sigma_7}{4^4\Delta_{37}\Delta_{17}\Delta_{57}} \text{sgn}(K_2^z) |\sigma_1,\sigma_2,\sigma_3,\sigma_4,\sigma_5,\sigma_6,\sigma_7\rangle~.\nonumber
\eea
\normalsize
So the eight terms coming from $\{abcd, abdc, bacd, badc\}$ cancel out, and the same is true for their inverse processes $\{dcba, cdba, dcab, cdab\}$. Next come the processes:
\small
\bea
H_{\text{eff}}^{(B,cbda)}  |\sigma_1,\sigma_2,\sigma_3,\sigma_4,\sigma_5,\sigma_6,\sigma_7\rangle&=& 
 =-\frac{\lambda\sigma_1\sigma_7}{4^4\Delta_{13}\Delta_{35}\Delta_{57}} \text{sgn}(K_2^z) |\sigma_1,\sigma_2,\sigma_3,\sigma_4,\sigma_5,\sigma_6,\sigma_7\rangle~,\nonumber\\
H_{\text{eff}}^{(B,cbad)}  |\sigma_1,\sigma_2,\sigma_3,\sigma_4,\sigma_5,\sigma_6,\sigma_7\rangle&=& 
=+\frac{\lambda\sigma_1\sigma_7}{4^4\Delta_{15}\Delta_{35}\Delta_{57}} \text{sgn}(K_2^z) |\sigma_1,\sigma_2,\sigma_3,\sigma_4,\sigma_5,\sigma_6,\sigma_7\rangle~,\nonumber
\eea
\normalsize
and similarly $H_{\text{eff}}^{(B,bcad)}\!=\!-H_{\text{eff}}^{(B,cbad)}$, and $H_{\text{eff}}^{(B,bcda)}\!=\!-H_{\text{eff}}^{(B,cbda)}$. So the processes coming from $\{cbad, cbda, bcad, bcda\}$ cancel out, and the same is true for their inverse processes $\{dabc, adbc, dacb, adcb\}$.

The only finite contributions then come from the remaining eight processes: $\{acbd,\!acdb,\!bdac,\!bdca\}$ and their inverses $\{dbca,\!bdca,\!cadb,\!acdb\}$. 
Here $H_{\text{eff}}^{(B,acbd)}\!=\!H_{\text{eff}}^{(B,cabd)}\!=\!H_{\text{eff}}^{(B,acdb)}\!=\!H_{\text{eff}}^{(B,cadb)}$, so there is no cancellation. We have:
\small
\bea
\mc{H}_{\text{eff}}^{(B,dbca)} |\sigma_1,\sigma_2,\sigma_3,\sigma_4,\sigma_5,\sigma_6,\sigma_7\rangle
=-\frac{\lambda \sigma_1\sigma_7}{4^4\Delta_{13}\Delta_{1357}\Delta_{15}} \text{sgn}(K_{2z}) |\sigma_1,\sigma_2,\sigma_3,\sigma_4,\sigma_5,\sigma_6,\sigma_7\rangle~.
\nonumber
\eea
\normalsize
In total, the effective terms arising from the NNN perturbations $K_2^{x(y)}$ is
\small
\be
\boxed{\mc{H}_{\text{eff}}^{(B)}\!=\!8 \mc{H}_{\text{eff}}^{(B,dbca)} \!=\! J_{1} S_1^z S_7^z},~~~
\boxed{J_{1}  = \frac{(K_2^x K_2^y)^2}{8(|K_1^z|+2|K_2^z|)^2 (2|K_1^z|+3|K_2^z|)} \text{sgn}(K_2^z)}~.
\ee
\normalsize
For $K_1^z\!=\!0$, $J_{1}\!=\!\frac{(K_2^x K_2^y)^2}{24\cdot 4 (K_2^z)^3} \text{sgn}(K_2^z)$, in agreement with the result obtained by Jackeli and Avella~\cite{jackeli15SM} for the triangular lattice case.

\subsection{Effective terms arising from mixed $K_1^{x(y)}$ and $K_2^{x(y)}$ perturbations.}\label{Sec:iii} 
Finally, we consider the perturbations due to  mixed $K_1^{x(y)}$ and $K_2^{x(y)}$ terms. Figure~\ref{fig:EffJs}~(C-H) shows the six minimal loops that contribute to an effective coupling of the form $J_{2}S_1^zS_4^z$, between sites $1$ and $4$. In the following we define $\kappa\!=\!K_1^x K_1^y K_2^x K_2^y$, and introduce the excitation energies of various intermediate virtual states: 
\small
\bea
&&\Delta_{12}\!=\!\Delta_{16}\!=\!\Delta_{14}\!=\!\Delta_{46}\!=\!\Delta_{24}\!=\!-|K_1^z|-2|K_2^z|, \nonumber\\
&&\Delta_{26}\!=\!\Delta_{35}\!=\!-|K_1^z|-|K_2^z|,~~~
\Delta_{23}\!=\!\Delta_{56}\!=\!-2|K_2^z|,\nonumber\\
&&\Delta_{1246}\!=\!\Delta_{1345}\!=\!-2|K_1^z|-3|K_2^z|, ~~~\Delta_{1234}\!=\!\Delta_{1456}\!=\!-|K_1^z|-4|K_2^z|.  \nonumber
\eea
\normalsize
Let us discuss the different processes C-H of Fig.~\ref{fig:EffJs} one by one.

\subsubsection{C \& D processes}
The perturbation $V\!=\!C$ described by the loops of type C of Fig.~\ref{fig:EffJs} splits as 
\small
\be\label{eq:C}
C=C_a+C_b+C_c+C_d=K_1^y S_1^yS_2^y+K_2^y S_2^yS_4^y+K_2^x S_4^xS_6^x+K_1^x S_6^xS_1^x.
\ee   
\normalsize
Replacing (\ref{eq:C}) into (\ref{eq:Heff-1}), we get twenty four contributions. We have
\small
\bea
\mc{H}_{\text{eff}}^{(C,dcba)} |\sigma_1,\sigma_2,\sigma_3,\sigma_4,\sigma_5,\sigma_6\rangle  &=& 
\mc{H}_{\text{eff}}^{(C,abcd)} |\sigma_1,\sigma_2,\sigma_3,\sigma_4,\sigma_5,\sigma_6\rangle =
\frac{-\kappa \sigma_1 \sigma_4}{4^4{ \Delta_{12}}{ \Delta_{14}}{ \Delta_{16}}} |\sigma_1,\sigma_2,\sigma_3,\sigma_4,\sigma_5,\sigma_6\rangle ~.\nonumber
\eea
\normalsize
We also find $\mc{H}_{\text{eff}}^{(C,dcab)}\!=\!\mc{H}_{\text{eff}}^{(C,cdba)}\!=\!\mc{H}_{\text{eff}}^{(C,dcba)}$. So all eight  processes $\{dcba,\!dcab,\!cdba,\!cdab\}$ and $\{abcd,\!bacd,\!abdc,\!badc\}$ give the same contribution.
Next come the processes of the type
\small
\bea
\mc{H}_{\text{eff}}^{(C,dbca)} |\sigma_1,\sigma_2,\sigma_3,\sigma_4,\sigma_5,\sigma_6\rangle  &=&
\mc{H}_{\text{eff}}^{(C,acbd)}|\sigma_1,\sigma_2,\sigma_3,\sigma_4,\sigma_5,\sigma_6\rangle  =
\frac{\kappa\sigma_1 \sigma_4}{4^4 { \Delta_{12}}{\Delta_{1246}}{\Delta_{16}}} |\sigma_1,\sigma_2,\sigma_3,\sigma_4,\sigma_5,\sigma_6\rangle~,\nonumber
\eea
\normalsize
and $\mc{H}_{\text{eff}}^{(C,dbac)}\!=\!\mc{H}_{\text{eff}}^{(C,bdca)}\!=\!\mc{H}_{\text{eff}}^{(C,dbca)}$. So all 8 processes $\{dbca\!,dbac,\!bdca,\!bdac\}$ and $\{acbd\!,cabd,\!acdb,\!cadb\}$ give the same contribution.
Finally there are the processes of the type:
\small
\bea
\mc{H}_{\text{eff}}^{(C,cbda)} |\sigma_1,\sigma_2,\sigma_3,\sigma_4,\sigma_5,\sigma_6\rangle &=&
\mc{H}_{\text{eff}}^{(C,adbc)}|\sigma_1,\sigma_2,\sigma_3,\sigma_4,\sigma_5,\sigma_6\rangle 
=-\frac{\kappa\sigma_1\sigma_4}{4^4\Delta_{12}\Delta_{26}\Delta_{46}}|\sigma_1,\sigma_2,\sigma_3,\sigma_4,\sigma_5,\sigma_6\rangle ~. \nonumber
\eea
\normalsize
Here, however, $\mc{H}_{\text{eff}}^{(C,cbad)} \!=\!-\mc{H}_{\text{eff}}^{(C,cbda)}$, and similarly $\mc{H}_{\text{eff}}^{(C,bcda)}\!=\!-\mc{H}_{\text{eff}}^{(C,cbda)}$. As a result, the last eight  processes $\{cbda,\!bcda,\!cbad,\!bcad\}$ and $\{adbc,\!adcb,\!dabc,\!dacb\}$ cancel out. So the total contribution from the $C$ loops of Fig.~\ref{fig:EffJs}~(C) is:
\small
\bea
\mc{H}_{\text{eff}}^{(C)} &=& 8 \mc{H}_{\text{eff}}^{(C,abcd)}+8 \mc{H}_{\text{eff}}^{(C,dbca)}=
\frac{\kappa \left(\Delta_{12}-\Delta_{1246}\right) }{32 \Delta_{12}^3\Delta_{1246}} \sigma_1\sigma_4~,\nonumber
\eea
\normalsize
where $\Delta_{12}\!-\!\Delta_{1246}\!=\!|K_1^z|\!+\!|K_2^z|\!>\!0$. So the coupling is AFM.  

Finally, by symmetry, $\mc{H}_{\text{eff}}^{(D)}\!=\!\mc{H}_{\text{eff}}^{(C)}$.

\subsubsection{E \& F processes}
These processes give rise to an overall constant, so they can be ignored.

\subsubsection{G \& H processes}
Here the corresponding perturbation can be written as
\small
\be\label{eq:G}
G=G_a+G_b+G_c+G_d=K_1^y S_1^yS_2^y+K_2^y S_2^yS_4^y+K_1^x S_3^xS_4^x+K_2^x S_1^xS_3^x ~.
\ee   
\normalsize
We have
\small
\bea
\mc{H}_{\text{eff}}^{(G,dcba)}|\sigma_1,\sigma_2,\sigma_3,\sigma_4,\sigma_5,\sigma_6\rangle 
 &=& 
\mc{H}_{\text{eff}}^{(G,abcd)} |\sigma_1,\sigma_2,\sigma_3,\sigma_4,\sigma_5,\sigma_6\rangle  =
 \frac{-\kappa\sigma_1\sigma_4}{4^4\Delta_{12}^3}  |\sigma_1,\sigma_2,\sigma_3,\sigma_4,\sigma_5,\sigma_6\rangle~,\nonumber
\eea
\normalsize
where we used the relation ${\Delta_{13}}\!=\!{\Delta_{14}}\!=\!\Delta_{12}$. Similarly, we can also show that $\mc{H}_{\text{eff}}^{(G,dcba)}\!=\!\mc{H}_{\text{eff}}^{(G,cdba)}\!=\!\mc{H}_{\text{eff}}^{(G,dcab)}$, so the eight processes $\{dcba,\!dcab,\!cdba,\!cdab\}$ and $\{abcd,\!bacd,\!abdc,\!badc\}$ give the same contribution.
Next come the processes of the type:
\small
\bea
\mc{H}_{\text{eff}}^{(G,dbca)} |\sigma_1,\sigma_2,\sigma_3,\sigma_4,\sigma_5,\sigma_6\rangle &=& 
\mc{H}_{\text{eff}}^{(G,acbd)}  |\sigma_1,\sigma_2,\sigma_3,\sigma_4,\sigma_5,\sigma_6\rangle 
=
\frac{\kappa\sigma_1\sigma_4}{4^4{\Delta_{12}}{\Delta_{1234}}{ \Delta_{13}}}  |\sigma_1,\sigma_2,\sigma_3,\sigma_4,\sigma_5,\sigma_6\rangle ~.\nonumber
\eea
\normalsize
Again, $\mc{H}_{\text{eff}}^{(G,dbca)}\!=\!\mc{H}_{\text{eff}}^{(G,dbac)}\!=\!\mc{H}_{\text{eff}}^{(G,bdca)}$. So all eight processes $\{dbca,\!dbac,\!bdca,\!bdac\}$ and $\{acbd,\!cabd,\!acdb,\!cadb\}$ give the same contribution.
Finally there are the processes of the type:
\small
\bea
\mc{H}_{\text{eff}}^{(G,cbda)}  |\sigma_1,\sigma_2,\sigma_3,\sigma_4,\sigma_5,\sigma_6\rangle &=&  
\mc{H}_{\text{eff}}^{(G,adbc)}  |\sigma_1,\sigma_2,\sigma_3,\sigma_4,\sigma_5,\sigma_6\rangle
= 
-\frac{\kappa\sigma_1\sigma_4}{4^4{\Delta_{12}}{ \Delta_{23}}{\Delta_{34}}}  |\sigma_1,\sigma_2,\sigma_3,\sigma_4,\sigma_5,\sigma_6\rangle~.\nonumber
\eea
\normalsize
Similarly, $\mc{H}_{\text{eff}}^{(G,cbad)}\!=\!\mc{H}_{\text{eff}}^{(G,bcda)}\!=\!-\mc{H}_{\text{eff}}^{(G,cbda)}$. So here, $\{cbda,\!cbad,\!bcda,\!bcad\}$ and $\{adbc,\!dabc,\!adcb,\!dacb\}$ cancel out. 

Altogether 
\small
\bea
\mc{H}_{\text{eff}}^{(G)}=8\mc{H}_{\text{eff}}^{(G,dcba)}+8\mc{H}_{\text{eff}}^{(G,dbca)}
=\frac{\kappa\left( \Delta_{12}- \Delta_{1234}\right)}{32\Delta_{12}^3\Delta_{1234} } \sigma_1\sigma_4 ~,
\nonumber
\eea
\normalsize
where $\Delta_{12}-\Delta_{1234}\!=\!2|K_2^z|\!>\!0$. So $\mc{H}_{\text{eff}}^{(G)}$ is also AFM. Finally, by symmetry, $\mc{H}_{\text{eff}}^{(H)}\!=\!\mc{H}_{\text{eff}}^{(G)}$.

\subsubsection{Final result}
\small
\be
\boxed{\mc{H}_{\text{eff}}^{(C-H)}=2\mc{H}_{\text{eff}}^{(C)}+2\mc{H}_{\text{eff}}^{(G)}=J_{2} S_1^z S_4^z},~~~~~
\boxed{J_{2}=-\frac{\kappa}{4\Delta_{12}^3} \left[\frac{|K_1^z|+|K_2^z|}{2|K_1^z|+3|K_2^z|} +\frac{2|K_2^z|}{|K_1^z|+4|K_2^z|} \right]}~.
\ee
\normalsize



%

\end{widetext}
\end{document}